\documentclass[journal]{IEEEtran}

\usepackage{cite}
\usepackage{amsmath,amssymb,amsfonts}
\usepackage{algorithmic}
\usepackage{graphicx}
\usepackage{textcomp}
\usepackage{xcolor}
\usepackage{subfig}
\usepackage[2]{editing}

\usepackage[linesnumbered,ruled,vlined]{algorithm2e}
\def\BibTeX{{\rm B\kern-.05em{\sc i\kern-.025em b}\kern-.08em
    T\kern-.1667em\lower.7ex\hbox{E}\kern-.125emX}}
\begin{document}

\title{NOMA Assisted Downlink Power Allocation in Pinching Antenna Systems Using Convolutional Neural Network
\thanks{$^\dag$School of Physics, Engineering and Technology - University of York, UK.  $^{\dag\dag}$The School of Electrical and Electronic Engineering, University of Manchester, Manchester, UK. }}


\author{Saeed~Mohammadzadeh,$^{\dag}$ Kanapathippillai~Cumanan,$^{\dag}$, and~Zhiguo~Ding,$^{\dag\dag}$}

\maketitle

\begin{abstract}
In this paper, we consider a flexible-antenna architecture, referred to as a pinching-antenna (PA) system, in which multiple PAs realized by activating small dielectric particles along a dielectric waveguide are jointly employed to serve a single-antenna user. Our goal is to investigate the best antenna placement and power allocation optimization in PA-assisted non-orthogonal multiple access (NOMA) systems using a convolutional neural network (CNN). An optimization strategy is first developed to determine the PA locations that maximize achievable NOMA performance while satisfying physical and spatial constraints. The proposed method adopts a two-stage structure, combining a user-aware initialization with a gradient-based refinement, enabling near-optimal performance with significantly lower computational cost. Then, a max–min fairness formulation is introduced for power allocation to balance the power budget among users with varying channel strengths, solved efficiently via quasi-linear programming and bisection search. Finally, a CNN-based learning framework is employed to capture the nonlinear mapping between channel conditions and corresponding optimal power coefficients. The CNN framework can infer near-optimal power allocations for unseen network configurations without retraining, offering scalability and adaptability. Simulation results demonstrate that the proposed CNN-based NOMA approach for PA systems delivers enhanced performance in terms of sum rate and user fairness with reduced computational complexity.
\end{abstract}


\section{Introduction}
\IEEEPARstart{A}{ddressing} the increasing connectivity demands of user equipments across diverse indoor environments poses a major challenge for next-generation wireless network design. In this evolving landscape, wireless systems must ensure robust downlink performance to support emerging applications such as high-definition video streaming, augmented reality, and real-time interactive services. These applications demand high data rates and low latency to guarantee seamless user experiences and efficient edge-network operation \cite{evgenidis2024multiple}. Although centralized multiple-input multiple-output (MIMO) technologies have demonstrated strong capabilities in mitigating small-scale fading and enhancing throughput, their performance remains limited by large-scale path loss, particularly in dense or heterogeneous deployment scenarios \cite{zhang2022beam}. Consequently, improving downlink efficiency continues to be a fundamental challenge in next-generation wireless systems.

Recent advancements in flexible-antenna technologies, such as movable antennas \cite{zhu2023modeling}, fluid antennas \cite{wong2021fluid}, and reconfigurable intelligent surfaces (RISs) \cite{huang2019reconfigurable}, have opened new opportunities for enhancing link reliability and coverage. However, each approach faces inherent limitations. For instance, RISs can mitigate line-of-sight (LoS) blockages by redirecting incident signals, but they suffer from double attenuation due to losses on both reflection paths \cite{tyrovolas2023zero}. Movable and fluid antennas, while reconfigurable and effective in constrained environments, are typically confined to base station (BS) installations and struggle to alleviate large-scale path loss, particularly at high frequencies where shorter wavelengths exacerbate signal degradation \cite{zhou2025exploiting}. Furthermore, under non-LoS conditions, path loss increases substantially, underscoring the need for maintaining strong LoS links. Existing flexible antenna solutions also face structural limitations, as their designs often restrict the addition or removal of antenna elements, reducing adaptability.
Given these challenges, pinching antenna (PA) systems have recently emerged as a promising alternative for indoor environments \cite{ding2024flexible, yang2025pinching, liu2025pinching}. A PA system comprises a waveguide integrated with multiple dielectric elements (e.g., plastic pinches) acting as antennas. Connected to the BS and installed along the ceiling edge, the electromagnetically isolated waveguide supports multiple PAs that enable flexible, scalable, and efficient downlink transmission \cite{suzuki2022pinching}.
\subsection{Related Works}
Several studies have investigated the performance of pinching antenna systems (PASS). In \cite{wang2025modeling}, a coupled-mode theory–based PASS model introduced equal and proportional power schemes and formulated a joint transmit and pinching beamforming optimization problem under continuous and discrete activations. Penalty-based and zero-forcing algorithms were proposed, achieving notable power savings with minimal performance loss.
Furthermore, \cite{tyrovolas2025performance} developed an analytical framework for the outage probability and average rate of PASS, accounting for both free-space path loss and waveguide attenuation, and characterized optimal antenna placement under realistic waveguide loss conditions. Similarly, \cite{hou2025performance} optimized the placement of antennas to maximize the effective channel gain, while \cite{ouyang2025array} examined the array gain under varying numbers and spacings of PAs along a single waveguide. Several works have extended PASS to integrated sensing and communication (ISAC) applications. The authors in \cite{qin2025joint} jointly optimized antenna positioning and transmit power to maximize total communication rate while meeting sensing and energy constraints. Similarly, \cite{zhang2025integrated} designed a PASS-based ISAC system to establish reliable LoS communication and sensing links, while \cite{ding2025pinching} analyzed channel reconfiguration and array flexibility from the Cramér–Rao lower bound perspective. These results demonstrate that PASS can significantly enhance ISAC performance by maintaining strong LoS paths and reconfigurable geometries.
The study in \cite{bereyhi2025downlink} investigated PASS-enabled multiuser MIMO downlink systems, formulating a joint digital precoding and antenna placement problem to maximize the weighted sum rate. A low-complexity iterative algorithm based on fractional programming was proposed for joint optimization. Similarly, \cite{mu2025pinching} explored long-spread waveguides with adjustable PAs for low-loss transmission and efficient free-space radiation, highlighting the potential of hybrid beamforming in practical PASS implementations.\\
\indent Placement and power optimization have been central themes in PASS research. The work in \cite{xie2025low} proposed a low-complexity closed-form solution for optimal PA placement to maximize the downlink sum rate, analyzing both orthogonal multiple access (OMA) and non-orthogonal multiple access (NOMA) scenarios. In \cite{tegos2024minimum}, uplink OMA transmission was optimized via successive convex approximation for efficient time resource allocation. Single-user rate maximization through antenna activation optimization was explored in \cite{xu2025rate}, while \cite{zeng2025energy} addressed joint antenna positioning and resource allocation for energy-efficient downlink transmission. Considering frequency-dependent attenuation, \cite{hu2025sum} jointly optimized precoding and PA placement to maximize sum rate. Moreover, integrating NOMA and OMA into PASS has gained attention; \cite{ding2025blockage} demonstrated that pinching-antenna systems effectively mitigate blockages and maintain LoS links, enhancing NOMA performance by suppressing co-channel interference. In \cite{mohammadzadeh2025efficient}, authors present a power allocation framework aimed at minimizing the total transmit power in a downlink NOMA system equipped with multiple PAs\\
Some works have explored optimization in PA-assisted NOMA systems. In \cite{zhou2025sum}, a bisection-based algorithm was developed for joint power allocation and antenna placement to maximize the sum rate. The study in \cite{wang2024antenna} proposed a dynamic PA selection scheme using a spherical wave model, while \cite{xu2025qos} optimized transmission for a two-user NOMA downlink. In \cite{fu2025power}, total power minimization across multiple dielectric waveguides was addressed via an iterative algorithm with guaranteed convergence. The authors in \cite{zhang2025uplink} formulated an uplink sum-rate maximization problem for multiuser PASS channels, solved under minimum mean square error decoding with successive interference cancellation (SIC). Authors in \cite{guo2025gpass} proposed a graph-based PASS framework that jointly learns pinching and transmit beamforming using a staged architecture of two sub-GNNs.
\subsection{Motivation and Contributions}
\indent In previously reported NOMA-based algorithms, user power allocation is typically fixed, overlooking the essential role of transmit power control in the operation of pinching antennas (PAs). Efficient power management is critical to guarantee reliable reception and minimize energy consumption. As more antennas are activated, transmit power must be shared among users, reducing the per-user power budget and potentially degrading performance. This limitation highlights the significance of NOMA, where superposition coding allows multiple users to share the same spectrum efficiently within the waveguide structure.
Antenna placement also exerts a major influence on achievable user data rates. However, most existing studies assume fixed or predefined antenna positions or employ exhaustive search-based optimization, which becomes impractical when the number of users or antennas varies. Any change in these parameters demands full re-simulation, restricting scalability and real-time adaptability. Consequently, the overall performance of a NOMA-based PA system depends on the joint optimization of antenna locations and power allocation. Fixed configurations lack the flexibility to adapt to user mobility and dynamic channel variations, making them unsuitable for large-scale or time-varying deployments.

Motivated by these considerations, a two-stage algorithm is designed to find near-optimal antenna positions that maximize the achievable NOMA sum-rate: (i) a user-centric initialization phase, where antennas are initially distributed across the user region and slightly shifted toward user-dense areas, and (ii) a gradient-based refinement phase, where antenna coordinates are iteratively updated to enhance the system sum-rate while maintaining constraints such as inter-antenna spacing and physical boundaries. Next, a power allocation framework is developed to maximize the minimum achievable data rate under quality-of-service (QoS) constraints, ensuring max–min fairness among users. The corresponding nonconvex optimization problem jointly captures the effects of large-scale path loss and the phase variations introduced by both in-waveguide and free-space propagation and is solved utilizing quasi-
linear programming. To further enhance adaptability and scalability, we introduce a learning-based framework that leverages deep convolutional neural networks (CNNs) to model the nonlinear relationships between user locations, antenna configurations, and optimal power allocation strategies. Once trained, the proposed CNN generalizes effectively to varying numbers of users and antennas without retraining, enabling a flexible and computationally efficient solution for dynamic network conditions. To the best of our knowledge, this is the first study integrating CNNs with waveguide-based PA systems in a NOMA context.
Simulation results confirm that the proposed NOMA-based framework significantly improves the achievable sum-rate compared to conventional OMA systems, particularly under optimized PA configurations. These results validate the potential of CNN-based learning as a powerful and scalable tool for joint antenna and power optimization in next-generation adaptive wireless communication systems.

The remainder of this paper is organized as follows. Section II introduces the system model of the pinching antenna and formulates the corresponding channel model. Section III presents the signal transmission schemes for pinching antenna systems under both OMA and NOMA frameworks. Section IV describes the proposed user-centric antenna placement strategy, formulates the max–min spectral efficiency optimization problem, and discusses the developed iterative antenna positioning algorithm while detailing the architecture of the proposed CNN-based power allocation framework, and evaluates the accuracy of the predicted outputs using multiple validation approaches. Section VI analyzes the computational complexity of the proposed method. Section VII provides numerical results and performance comparisons with existing benchmark schemes. Finally, Section VIII concludes the paper and summarizes the key findings.

\textit{Notations:}  The vector and matrix symbols are represented as lowercase and uppercase bold. $\mathbb{E} \{\cdot\}$ stands for the statistical expectation of random variables, and a circular symmetric complex Gaussian matrix with covariance $\mathbf{B}$ is denoted by $\mathcal{CN}(0,\mathbf{B})$. The symbol $\mathbb{R}$ is used for the real numbers. $(\cdot)^\mathrm{T}$, $(\cdot)^\mathrm{*}$, and $|\cdot |$ denote transpose, conjugate, and absolute value, respectively. We use the superscript $^\text{pin}$ to represent the variables or parameters associated with pinching antennas. The Euclidean norm of a matrix is denoted by $\|\cdot \|_2$.  The real and imaginary parts are presented as $\mathfrak{R}$ and $\mathfrak{J}$, respectively. 
\begin{figure}[t]
	\centering
		\includegraphics[width=0.9\linewidth]{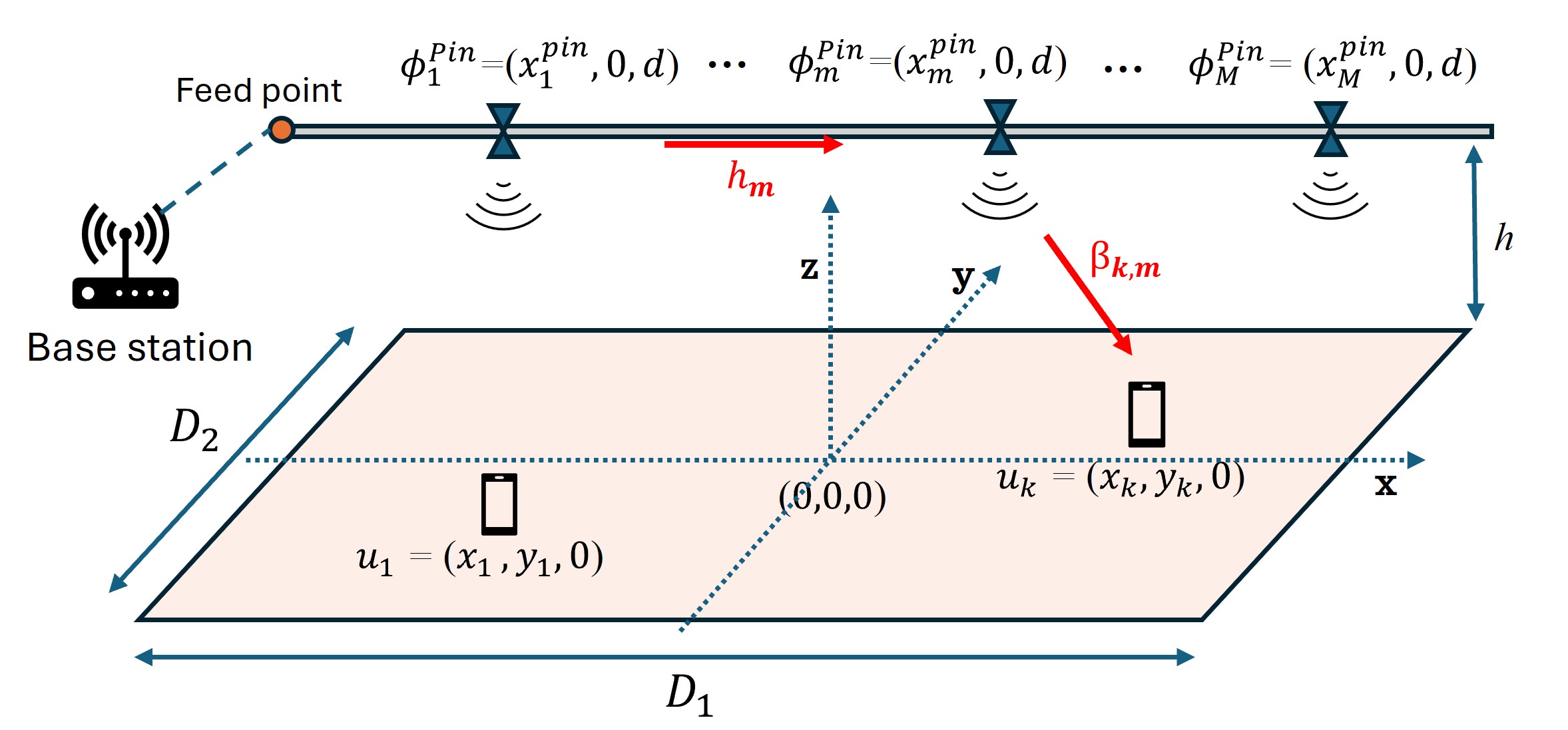}
	\caption{\small{Illustration of a network with a single waveguide. The $k$-th user is served and $\phi_m^\text{pin}$ pinching antennas is activated.}}
	\label{fig:fig1}
\end{figure}

\section{System Model}

We consider a downlink communication system shown in Fig.~\ref{fig:fig1} consisting of a BS that is connected to a single waveguide with $M$ PAs and $K$ single antenna users. Considering a three-dimensional coordination system where the users are uniformly distributed within an area $D_2 \times D_1$ with its center at $(0,0,0)$. Moreover, we assume that the devices are randomly deployed in a square area lying in the $x-y$ plane, and $u_k = (x_k,y_k,0)$ denotes the position of the $k$-th user \cite{ding2024flexible}. $x_k$ and $y_k$ follows a uniform distribution in $[-D_1/2 , D_1/2]$ and $[-D_2/2 , D_2/2]$, respectively. In the PA system, the waveguide is assumed to be placed parallel to the x-axis \cite{yang2025pinching}, where the height of the waveguide is denoted by $d$. Also, the location of the conventional antenna case is $\phi_0 = (0,0, d)$ while the location of the $m$-th PA is expressed as $\phi_m^\text{pin} = (x_m^\text{pin},0,d) $.\\
\indent Given that the BS is equipped with multiple antennas, serving \( K \) users concurrently becomes a natural approach. Let \( s \) denote the signal passed through the waveguide. By treating these antennas as elements of a conventional linear array, the system can be effectively analyzed and optimized using well-established linear array processing methodologies, and the received signal at $k$-th user can be expressed as follows: 
\begin{align} \label{original received}
    r_k = \sum_{m=1}^M g_{k.m} \: s + n_k,
\end{align}
where $n_k$ denotes the additive white Gaussian noise at the BS with zero mean and variance $\sigma^2$ (i.e., $n_k \sim \mathcal{CN}(0, \sigma^2)$), and $g_{k,m} = h_m \: \beta_{k,m}$, is the channel between the BS to the $k$-th user, which is the combination of the signal arriving at the  $m$-th PA from the BS, undergoing a guided propagation phase shift, $h_m$ and the channel between the PAs and the $k$-th user, $\beta_{k,m}$. We assume perfect channel state information (CSI) is available at the BS. \\
Since all $M$ PAs are aligned along a common waveguide, the signal radiated by each antenna can be regarded as a phase-shifted version of the original signal provided by the BS at the waveguide's feed point (\textit{feed point refers to the location that the signal is introduced into the waveguide}) \cite{pozar1998microwave}. Therefore, the effective channel along the waveguide can be modeled as:
\begin{align}\label{waveguide channel}
   h_m =  \sqrt{\frac{P}{M}} e^{-j \frac{2 \pi}{\lambda_g} | {\phi}_0^{\textbf{pin}} - {\phi}_m^{\text{pin}} |},
\end{align}
where $| \cdot |$ and $P$ denote the absolute value and total transmit power, respectively. To enable a tractable and insightful performance analysis, it is assumed that the total transmit power is uniformly distributed across the $M$ activated PAs. $\lambda_g = \lambda/ \kappa$ defines the wavelength in a dielectric waveguide, $\kappa$ is the effective refractive index of the dielectric waveguide, $ \lambda$ presents the carrier frequency \cite{pozar1992microstrip}, and $\phi_0^{\text{pin}}$ stands for the position of the feed point of the waveguide.\\
\indent The free-space propagation from the $m$-th antenna to the $k$-th user is modeled as:
\begin{align} \label{pathloss channel}
   \beta_{k.m} = \frac{\sqrt{\eta} \, e^{-j\scriptstyle \frac{2 \pi}{\lambda} | u_k - {\phi}_m^{\text{pin}} |}}{ | u_k - {\phi}_m^{\text{pin}} |},
\end{align}
where $ \eta = (c/4 \pi f_c)^2$, with $f_c$, $c$, and $\lambda$ representing the carrier frequency, speed of light, and the wavelength in free space, respectively. 
Based on these definitions, the channel for the $k$-th user is defined as
 \begin{align} \label{channel_g}
     g_k &= \sum_{m=1}^M g_{k,m}\\ \nonumber &= \sqrt{\dfrac{P}{M}} \sum_{m=1}^M \frac{\sqrt{\eta} e^{-j 2 \pi (\scriptstyle \frac{1}{\lambda} | u_k - \phi_m^{\text{pin}} | - \scriptstyle \frac{1}{\lambda_g} | \phi_0^{\text{pin}}-\phi_m^{\text{pin}}  |)} }{ | u_k -  \phi_m ^{\text{pin}}|},
 \end{align}
By replacing \eqref{channel_g} into \eqref{original received}, the received signal by $k$-th user is expressed as
\begin{align} \label{Recived Signal}
    r_k = \sqrt{\dfrac{P}{M}} \Big(  \sum_{m=1}^M \frac{\sqrt{\eta} e^{-j 2 \pi (\scriptstyle \frac{1}{\lambda} | u_k - \phi_m^{\text{pin}} | - \scriptstyle \frac{1}{\lambda_g} | \phi_0^{\text{pin}}-\phi_m^{\text{pin}}  |)} }{ | u_k -  \phi_m ^{\text{pin}}|}  \Big) s + n_k.
\end{align}

The received signal by the $k$-th user presented in \eqref{Recived Signal} highlights a distinctive characteristic of PAs. Unlike conventional antennas, PAs offer additional degrees of freedom, as both the large-scale path loss, \( | u_k - \phi_m^{\text{pin}}| \), and the phase shift term, \( \frac{2 \pi }{\lambda_g} | \phi_0^{\text{pin}} - \phi_m^{\text{pin}}| \), can be dynamically reconfigured by adjusting the physical placement of the PAs along the waveguide. 

\section{Signal Transmission for NOMA-Based PA Systems}
Unlike conventional MIMO systems, where each antenna transmits an independent signal, pinching-antenna systems require all antennas on the same waveguide to be excited by a common signal. This fundamental difference imposes a constraint whereby the transmitted signal must be a superposition of all user signals intended for transmission. Consequently, this architecture necessitates the use of NOMA to support multi-user communication. Through superposition coding at the BS and SIC at the receivers, multiple users can be served simultaneously in the downlink. This design efficiently leverages the passive nature of PAs and the shared waveguide infrastructure while maintaining the capability to handle multiple users concurrently.\\
The superimposed signal transmitted from the BS through the wavelength is:
\begin{align}
    s = \sum_{k=1}^K \sqrt{q}_k \: s_k,  \quad \text{and} \quad \mathbb{E} \{|s_k|^2 \} = 1.
\end{align}
where $\mathbb{E} \{\cdot\}$ stands for the statistical expectation of random variables, $s_k$ and $q_k$ are the signal and the power allocation coefficient for the $k$-th user, respectively. Since all $M$ pinching antennas on the waveguide are activated and the total transmit power $P$ is equally distributed among them, each antenna transmits with power $ p^{\text{pin}} = \frac{P}{M}$ while sending the same signal $s$. Accordingly, the received signal in \eqref{original received} can be reformulated as follows:
\begin{align}
    r_k = \sqrt{q}_k \: p^{\text{pin}} g_k^*\: s_k + p^{\text{pin}} \sum_{j=1,j \neq k}^K \sqrt{q}_j g_k^* s_j + n_k,
\end{align}
where $(\cdot)^*$ denote conjugate numbers. 
In the NOMA-based system, users decode their messages based on their channel strengths. The users are ordered such that their channel gains satisfy $|g_1|^2 \le |g_2|^2 \le \cdots \le |g_K|^2$, where user 1 has the weakest and user $K$ the strongest channel. SIC is applied at the receivers to enable this decoding strategy. Specifically, each user $k$ is able to decode and subtract the signals of users with weaker channels (i.e., users $1$ to $k-1$), while treating the signals of users with stronger channels (i.e., users $k+1$ to $K$) as interference \cite{S5}. Consequently, for a user $l \in \{k, k+1, \ldots, K\}$ attempting to detect the signal of user $k$, the received signal is given by:
\begin{equation}
r_{l,k} = \sqrt{q_k} \: p^{\text{pin}} g_{l}^* s_k + p^{\text{pin}} \sum_{j = k+1}^{K} \sqrt{q_j} g_l^* s_j + n_l,\label{idc1}
\end{equation}
Hence, user $l$ decodes the message intended for user $k$ with the following the signal-to-interference-and-noise ratio (SINR):
\begin{align}
    \mathrm{SINR}_{l,k} = \frac{|g_l|^2  \: q_k}{\sum_{j=k+1}^K |g_l|^2  \: q_j + \sigma^2_\text{pin}},
\end{align}
where $\sigma^2_\text{pin} = \sigma^2/p^{\text{pin}}$. According to the principle of power-domain NOMA, the user $k$ will decode $l$-th user's signal, $1 \leq l\leq k-1$, before decoding its own signal, which
means that the data rate of the $k$-th user's signal is given by \cite{ding2014performance} 
\begin{align}
    \mathrm{R}_k = \min \{\mathrm{R}_{l,k} \},  \quad  \forall l =k,k+1, \cdots, K,
\end{align}
where $\mathrm{R}_{l,k}$ denotes the data rate at which the $l$-th user can decode the $k$-th user's signal, and is defined as
\begin{align}
    \mathrm{R}_{l,k} = \log_2 \big(1+ \mathrm{SINR}_{l,k}  \big).
\end{align}
\section{Proposed User-Centric Antenna Placement and Power Allocation Framework}

This section presents the proposed optimization framework for PA-assisted NOMA systems. The objective is to optimize the spatial positions of the antennas and the transmit power allocation to maximize system performance while maintaining fairness and practical feasibility. 
\subsection{Proposed User-Centric Antenna Placement Method}
In this subsection, a user-centric antenna placement method is developed to determine the optimal positions of PAs along the waveguide, aiming to maximize the achievable NOMA sum-rate. The proposed method employs a two-stage optimization structure that combines a user-aware initialization with a gradient-based refinement to provide near-optimal performance at low computational cost. We then discuss that the proposed method maintains robust performance across a wide range of tuning parameter values $\alpha$, highlighting its ability to effectively balance uniform coverage and user-centric adaptation.

We consider the problem of determining the optimal positions of the $M$ pinching antenna elements along a one-dimensional waveguide to maximize the achievable NOMA sum-rate for $K$ users. The users are located at positions 
$u_k = (x_k,y_k,0)$ while all antennas are constrained to lie along the $x$-axis at a fixed height $h$. The $x$-positions of the antennas are collected into the vector
\begin{align}
    \mathbf{a} = [a_1,a_2,\ldots,a_M]^\top, \quad a_m \in \mathcal{A}=[a_{\min},a_{\max}],
\end{align}
where $\mathcal{A}$ denotes the feasible interval for antenna placements, e.g., the span of the user coordinates or the physical limits of the waveguide.  

\subsubsection{Optimization Problem}
The optimization objective is to maximize the achievable NOMA sum-rate (SR) under optimized power allocation:
\begin{align}
   \begin{array}{cl}
\max_{\mathbf{a}} & \mathrm{SR}(\mathbf{a} ; \mathbf{U}) \\
\text { s.t. } & a_{\min } \leq a_m \leq a_{\max }, \quad \forall m=1, \ldots, M, \\
& \left|a_i-a_j\right| \geq \tilde{\Delta}, \quad \forall i \neq j .
\end{array} 
\end{align}
where  $\tilde{\Delta}$ is the guard distance to avoid antenna coupling and  $\mathbf{U}=[u_1,u_2,\ldots,u_K]$ collects all user locations. The achievable sum-rate is defined as
\begin{align}
    \text{SR}(\mathbf{a};\mathbf{U}) = \sum_{k=1}^K R_k(\mathbf{a};\mathbf{U}),
\end{align}
with the achievable rate of user $k$ expressed as
\begin{equation}
    R_k(\mathbf{a};\mathbf{U}) = \log_2\!\left(1 + \frac{g_k(\mathbf{a})\,q_k}{\sigma^2 + g_k(\mathbf{a})\sum_{j>k} q_j}\right),
\end{equation}
For notational simplicity, $\text{SR}(\mathbf{a}; \mathbf{U})$ is denoted as $\text{SR}(\mathbf{a})$ throughout the remainder of the manuscript. To solve the non-convex problem of maximizing $\text{SR}(\mathbf{a})$, we adopt a two-stage approach consisting of an initialization stage followed by a gradient-based refinement stage.

\textit{Stage I: User-Centric Initialization.}
We first compute the span of the user $x$-coordinates:
\begin{align}
x_{\min} = \min_{k} x_k, 
\qquad 
x_{\max} = \max_{k} x_k.
\end{align}
The antennas are initially placed at equally spaced positions along this interval:
\begin{equation}
    a_m^{(0)} = x_{\min} + \frac{m-1}{M-1}\,(x_{\max}-x_{\min}), 
    \qquad m=1,\ldots,M.
\end{equation}
This ensures a uniform coverage of the user region.  
Next, to bias the placement toward the actual user distribution, each antenna is nudged toward its nearest user. Specifically, define
\begin{equation}
    k_m = \arg\min_{k} \big|x_k - a_m^{(0)}\big|,
\end{equation}
as the index of the closest user to antenna $m$. Then the initialization is refined via
\begin{equation}
    a_m^{(0)} \;\leftarrow\; (1-\alpha)\,a_m^{(0)} + \alpha\,x_{k_m}, 
    \qquad 0<\alpha<1,
\end{equation}
where $\alpha$ is a tuning parameter that balances uniform coverage and user-centric placement.  

\textit{Stage II: Gradient-Based Refinement.}
Starting from $\mathbf{a}^{(0)}$, the antenna positions are refined using gradient ascent. Let $\mathbf{a}^{(i)}$ denote the antenna positions at iteration $i$. The gradient of the sum-rate with respect to each antenna position $a_m$ is approximated numerically by finite differences. For a small perturbation $\delta>0$, we compute
\begin{equation}
    \frac{\partial \text{SR}}{\partial a_m}\bigg|_{\mathbf{a}^{(i)}} 
    \approx \frac{\text{SR}(\mathbf{a}^{(i)} + \delta \mathbf{e}_m) 
    - \text{SR}(\mathbf{a}^{(i)})}{\delta},
\end{equation}
where $\mathbf{e}_m$ is the $m$-th standard basis vector in $\mathbb{R}^M$.  
The antenna positions are then updated in the ascent direction:
\begin{equation}
    a_m^{(i+1)} = a_m^{(i)} + \eta_{\text{step}} 
    \cdot \frac{\partial \text{SR}}{\partial a_m}\bigg|_{\mathbf{a}^{(i)}},
\end{equation}
where $\eta_{\text{step}}$ is a step-size parameter. To guarantee feasibility, each update is projected back onto the feasible sets $\mathcal{F}$ defined by spacing and boundary constraints:
\begin{equation}
    a_{m+1}^{(i+1)}-a_m^{(i+1)} \geq \tilde{\Delta}, \quad a_{\min } \leq a_m^{(i+1)} \leq a_{\max }
\end{equation}
Projection is implemented by forward clamping, ensuring that
\begin{equation}
    a_m^{(i+1)}=\operatorname{clip}\left(a_m^{(i+1)}, a_{m-1}^{(i+1)}+\tilde{\Delta}, a_{\max }-(M-m) \tilde{\Delta}\right),
\end{equation}
where $ \text{clip} (x,l,u)=\max(l,\min(x,u))$. The iterative refinement continues until the relative improvement in the objective becomes negligible, i.e.,
\begin{equation}
    \big|\text{SR}(\mathbf{a}^{(i+1)}) - \text{SR}(\mathbf{a}^{(i)})\big| < \varepsilon,
\end{equation}
where $\varepsilon$ is a prescribed convergence tolerance. The converged antenna positions are denoted as
\begin{equation}
    \mathbf{a}^\star = [a_1^\star,a_2^\star,\ldots,a_M^\star]^\top,
\end{equation}
and serve as the optimized pinching antenna configuration that maximizes the achievable NOMA sum-rate.  The proposed idea is summarized in Algorithm 1.

\begin{algorithm}[t]
\caption{Proposed Pinching Antenna Position Optimization}
\label{alg:proposed}
\KwIn{User locations $\mathbf{U}=\{u_k=(x_k,y_k,0)\}_{k=1}^K$, 
number of antennas $M$, feasible range $\mathcal{A}=[a_{\min},a_{\max}]$, 
minimum spacing $\tilde{\Delta}$, step size $\eta_{\text{step}}$, 
tolerance $\varepsilon$, perturbation $\delta$, initialization weight $\alpha$.}
\KwOut{Optimized antenna positions $\mathbf{a}^\star$.}

\BlankLine
\textbf{Feasible set:}
\begin{align*}
\mathcal{F} = \{\mathbf{a}\in\mathbb{R}^M &: a_{\min}\le a_1\le\cdots\le a_M\le a_{\max},\\
&a_{m+1}-a_m\ge\tilde{\Delta},~\forall m\},
\end{align*}
feasible if $a_{\max}-a_{\min}\ge(M-1)\tilde{\Delta}$.

\BlankLine
\textbf{Stage I: User-Centric Initialization}\\
Compute user span $x_{\min}=\min_{k} x_k$, $x_{\max}=\max_{k} x_k$.\\
Initialize evenly spaced antennas:
\[
a_m^{(0)} = x_{\min} + \frac{m-1}{M-1}(x_{\max}-x_{\min}),~m=1,\ldots,M.
\]
For each $m$, find nearest user 
$k_m=\arg\min_k |x_k-a_m^{(0)}|$, 
and refine:
\[
a_m^{(0)} \leftarrow (1-\alpha)a_m^{(0)}+\alpha x_{k_m}.
\]

\BlankLine
\textbf{Stage II: Gradient-Based Refinement}\\
Set iteration $i=0$ and evaluate initial $\text{SR}(\mathbf{a}^{(0)})$.\\
\While{$|\mathrm{SR}(\mathbf{a}^{(i+1)})-\mathrm{SR}(\mathbf{a}^{(i)})|>\varepsilon$}{
\For{$m=1,\ldots,M$}{
\smallskip
(a) Perturb antenna $m$ by $\delta$: 
$\mathbf{a}^{(i)}+\delta\mathbf{e}_m$.\\
(b) Approximate gradient:
\[
\frac{\partial \mathrm{SR}}{\partial a_m} \approx 
\frac{\mathrm{SR}(\mathbf{a}^{(i)}+\delta\mathbf{e}_m)
-\mathrm{SR}(\mathbf{a}^{(i)})}{\delta}.
\]
}
(c) Update: $a_m^{(i+1)} = a_m^{(i)} + 
\eta_{\text{step}}\frac{\partial\text{SR}}{\partial a_m}$.\\
(d) Project onto feasible region:
\begin{align*}
a_1^{(i+1)} &\!\!\leftarrow\!
\text{clip}(a_1^{(i+1)},a_{\min},a_{\max}\!-\!(M\!-\!1)\tilde{\Delta}),\\
a_m^{(i+1)} &\!\!\leftarrow\!
\text{clip}(a_m^{(i+1)},a_{m-1}^{(i+1)}\!+\!\tilde{\Delta},a_{\max}\!-\!(M\!-\!m)\tilde{\Delta}),
\end{align*}
where $\text{clip}(x,l,u)=\max(l,\min(x,u))$.\\
(e) Update $\mathrm{SR}(\mathbf{a}^{(i+1)})$, set $i\!\leftarrow\! i\!+\!1$.
}
\BlankLine
Return $\mathbf{a}^\star=\mathbf{a}^{(i)}$.
\end{algorithm}

\begin{figure}[!]
	\centering
		\includegraphics[width=0.45\textwidth]{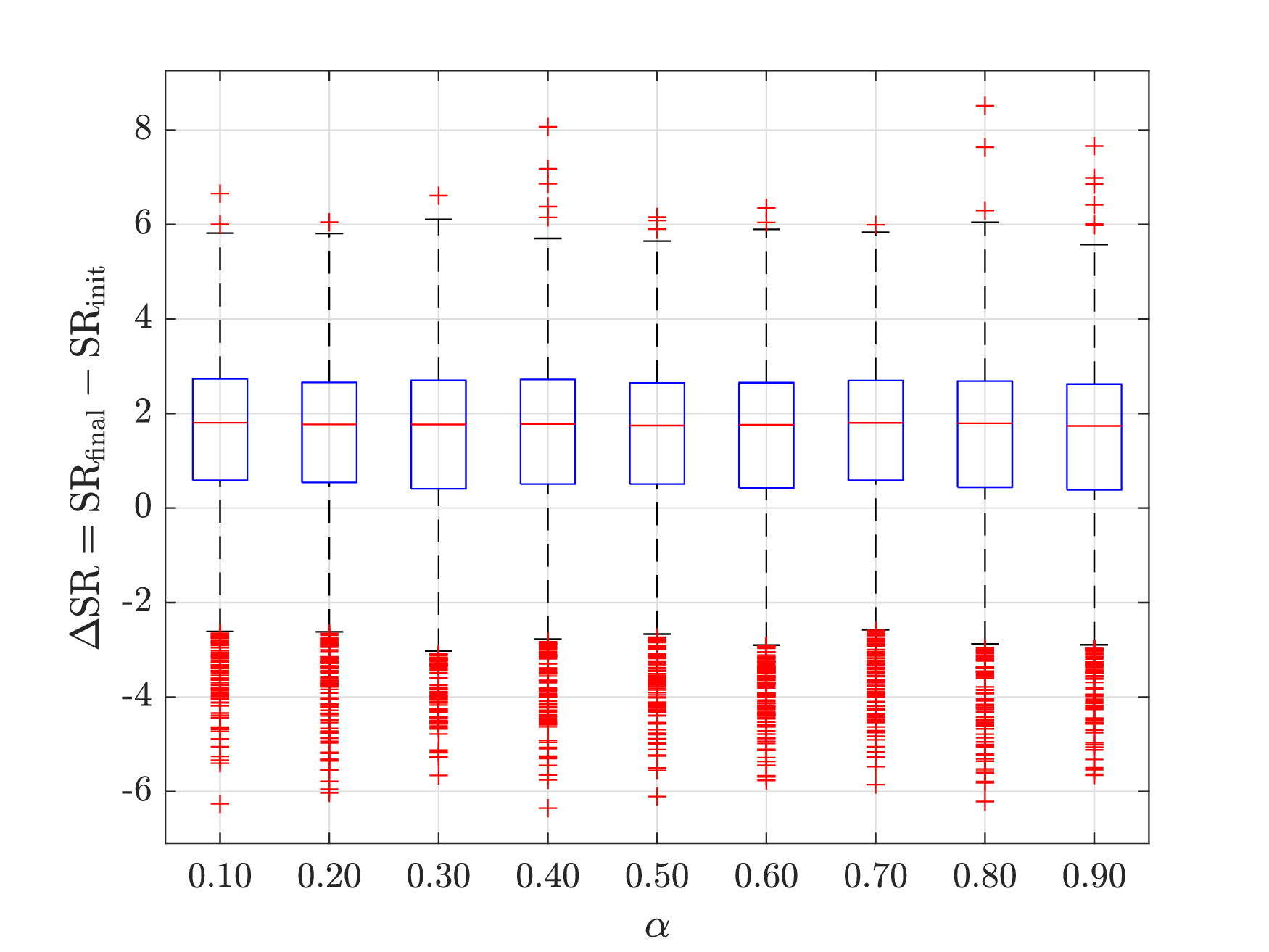}
        \vspace{-0.5em}
	\caption{Statistical visualization that summarizes the distribution of $ \Delta \text{SR}$ for 2000 random user layouts with respect to each $\alpha$. \textcolor{blue}{Blue box} spans from the 25th percentile to the 75th percentile of the data. This is the “interquartile range” (IQR). \textcolor{red}{Red line inside the blue box} is the median of all $\Delta$SR values. If this line is above zero, it means at least half of the experiments had improvement. \textbf{Black} dashed whiskers cover the most extreme values that are not classified as outliers (often defined as within 1.5 × IQR from the box edges). It shows the overall spread of the results. \textcolor{red}{+ lines} are Outliers and define high/low runs.} 
	\label{Boxplot}
\end{figure}
\subsubsection{Discussion}
In the proposed method, the parameter $\alpha$ is the tuning parameter for balancing uniform coverage and user-centric bias. In the following, we demonstrate that our method consistently improves performance independent of $\alpha$. We performed a robustness study for $M=5$, $K=4$, and the transmit power is 10 dBm over $\alpha \in (0,1)$ with 2000 random user layouts. For each $\alpha$, we recorded the sum-rate after initialization ($\textit{Stage I} $), $\text{SR}_\text{init}$, and after gradient refinement ($\textit{Stage II}$), $\text{SR}_\text{final}$ and calculate $\Delta  \text{SR} = \text{SR}_\text{final} - \text{SR}_\text{init}$. We want to demonstrate that the gradient refinement consistently improves SR across the full range of initialization $\alpha$. To provide a statistically robust evaluation, we employ boxplots to present the whole distribution of improvements across multiple Monte Carlo trials. 
Fig.~\ref{Boxplot} visualizes the median, interquartile range (IQR), whiskers, and outliers. Specifically, the central line inside the box corresponds to the median performance, the box itself spans the interquartile range (25th–75th percentiles), and the whiskers extend to cover typical variations within 1.5 times the IQR. Outliers beyond this range are plotted individually and represent rare realizations with significantly different performance.

This representation offers two main insights. First, it highlights the statistical robustness of the proposed scheme by showing that the median performance consistently improves after refinement. In other words, regardless of the choice of $\alpha$, the refinement stage enhances the sum-rate in the majority of cases. Second, the fact that all median values remain strictly above zero across the entire range of $\alpha$ indicates that the method provides a consistent and reliable gain, independent of the initialization parameter. Overall, the distribution of SR improvements demonstrates that gradient refinement yields a positive and stable improvement in performance for tested initialization settings.

\subsection{Max-Min spectral efficiency fairness}

In this subsection, we present the optimization of the power allocation by introducing a max–min fairness formulation. This approach ensures power budget distribution among users with different channel strengths through a quasi-linear programming formulation solved by bisection search. Then, we evaluate the efficiency and near-optimality of the proposed framework compared to a brute-force benchmark, demonstrating the effectiveness and scalability of the overall approach.
\subsubsection{Optimization Problem}
We consider the max–min fairness optimization problem in a downlink NOMA system. The max–min fairness criterion is particularly well-suited to the shared-waveguide architecture of PA systems. 
Unlike conventional MIMO, where independent antennas provide spatial degrees of freedom, PAs exhibit strong signal superposition within the waveguide, leading to tightly coupled interference among users. 
Under such conditions, proportional-fair or sum-rate objectives often prioritize users with favorable channels, leaving weaker users at a disadvantage. 
By contrast, the max–min fairness criterion guarantees each user a minimum SINR, thereby ensuring balanced service quality despite the inherent interference coupling of the waveguide. 
This observation is consistent with our numerical results, which demonstrate that max–min fairness achieves more equitable performance than alternative utility-based objectives. 
Moreover, the flexible rate and power allocation capabilities of NOMA make it particularly suitable for enforcing such fairness, as the max–min formulation explicitly maximizes the minimum achievable data rate across all users under a total transmit power constraint. This leads to the formulation of the following max-min fairness problem:
\begin{align} \label{sm23}
 &\max_{\mathbf{q} }\; \min_{k\in \{1,\cdots,K\}} \mathrm{R}_{l,k }(\mathbf{q})
\nonumber \\ & \; \text{s.t.  } \: \mathbf{1}^T \mathbf{q}   \le P, 
\\ & \; \;\;\;\; \; \mathbf{q} \ge \mathbf{0} , \nonumber
\end{align}
where $\mathbf{q} = [q_1, \cdots, q_K]^T$ is the set of transmit power coefficients for all the $K$ users. 
Since the spectral efficiency of user $k$ is given by $\log_2 \big(1+ \mathrm{SINR}_{l,k}(\mathbf{q}) \big)$, which is a monotonically increasing function of the effective $\mathrm{SINR}$, maximizing the minimum spectral efficiency is equivalent to maximizing the minimum effective $\mathrm{SINR}$ across all users. Therefore, the original problem in \eqref{sm23} can be reformulated as:
\begin{subequations}\label{sm24}
\begin{align} 
 &\max_{\mathbf{q} }\; \min_{k\in \{1,\cdots,K\}} \mathrm{SINR}_{l,k }(\mathbf{q})
\nonumber \\ & \; \text{s.t.  } \: \mathbf{1}^T \mathbf{q}   \le P, \label{total power}
\\ & \; \;\;\;\; \; \mathbf{q} \ge \mathbf{0} ,\label{qge}  
\end{align}
\end{subequations}
where $\mathbf{1} = [1 \cdots 1] \in \mathbb{R}^K$. The $P$ represents the total transmit power available at the BS. The constraint \eqref{total power} ensures that the total power of the users is below the available power threshold. As shown in Appendix~A, the combined effects of large-scale path loss and phase shifts (arising from in-waveguide and free-space propagation) render the SINR expression non-convex. Consequently, the max–min formulation in \eqref{sm24} is inherently non-convex, which makes the optimization problem challenging to solve directly.
To address this, we utilize quasi-linear programming and introduce a variable $t$, representing the minimum SINR for all users, which we then maximize. Consequently, \eqref{sm24} can be written as follows
\begin{align} \label{sm25}
 &\max_{\mathbf{q},\: t \ge 0 }\; t \nonumber \\ 
 & \; \text{s.t.  } \: \mathrm{SINR}_{l,k}(\mathbf{q}) \ge t, \quad \forall k,\\
 & \eqref{total power}-\eqref{qge},  \nonumber
\end{align}
where $t \triangleq  \min_k \mathrm{SINR}_{l,k} $. The optimal value of the problem in \eqref{sm25} can be efficiently determined using a bisection search over the parameter $t$ as shown in Algorithm 2. 
At each iteration, we consider a value $t = t^*$ and need to solve the following feasibility problem:
\begin{align} \label{find q}
 & \text{find} \quad \mathbf{q}  \nonumber \\ 
 & \; \text{s.t.  } \: \mathrm{SINR}_{l,k}(\mathbf{q}) \ge t^*, \quad \forall k,\\
 & \eqref{total power}-\eqref{qge},  \nonumber
\end{align}
The goal of the feasibility problem in \eqref{find q} is to find a vector $\mathbf{q}$ satisfying all constraints. Following \cite{demir2021foundations}, it can equivalently be reformulated to include minimization of the total transmit power $\sum_{k=1}^K q_k$ while preserving the original constraints. Based on this reformulation, Algorithm 2 is developed, with its subproblem defined in \eqref{inside algo}.

\begin{algorithm}[t]
\caption{Max-Min problem in \eqref{sm25} based on bisection search}
\DontPrintSemicolon
\KwIn{Tolerance $\zeta > 0$, maximum total power $P_{\text{total}}$}
\KwOut{Optimal power vector $\mathbf{q}^{\text{opt}}$, $t^{\text{opt}}$}

Initialize: $t_{\text{min}} \gets 0$, \: $t_{\text{max}} \gets \min_{k \in \{1, \ldots, K\}} (\sigma^2 /|g_k|^2)$ \\
$\mathbf{q}^{\text{opt}} \gets \mathbf{0}_K$, $t^{\text{opt}} \gets 0$ \\

\While{$t_{\text{max}} - t_{\text{min}} > \zeta$}{
    $t^{\text{*}} \gets \frac{t_{\text{min}} + t_{\text{max}}}{2}$ \;

    Solve the following problem with $t = t^{\text{*}}$:
    \begin{equation} \label{inside algo}
        \begin{aligned}
            &\underset{\mathbf{q} \geq \mathbf{0}_K}{\min} \quad \sum_{k=1}^K q_k \\
            &\text{s.t.} \quad \mathrm{SINR}_{l,k}(\mathbf{q}) \geq t^{\text{*}}, \quad \forall k \\
            & \quad \: \; \;\;\;\; \; \mathbf{q} \ge \mathbf{0}
        \end{aligned}
    \end{equation}

    \eIf{the problem is feasible}{
        $t_{\text{min}} \gets t^{\text{*}}$ \;
        $\mathbf{q}^{\text{opt}} \gets \mathbf{q}$ (optimal solution found) \;
    }{
        $t_{\text{max}} \gets t^{\text{*}}$ \;
    }
}
$t^{\text{opt}} \gets \min_{k} \mathrm{SINR}_{l,k}(\mathbf{q}^{\text{opt}})$ \;
\Return $\mathbf{q}^{\text{opt}}, t^{\text{opt}}$
\end{algorithm}
\subsubsection{Validation and Robustness}
To evaluate the effectiveness of the proposed iterative algorithm, we compare its performance with an exhaustive brute-force (BF) search that provides the true optimal solution over a discretized search space. Let $F(\mathbf{a})$ denote the objective function (sum rate) for antenna placement $\mathbf{a}$, and let $\mathbf{a}_{\mathrm{BF}}$ and $\mathbf{a}_{\mathrm{it}}$ be the solutions obtained by brute-force and the iterative algorithm, respectively. The deviation from optimality is measured by the relative gap
\begin{align}
   \mathrm{gap}_t = \frac{F(\mathbf{a}_{\mathrm{BF},t}) - F(\mathbf{a}_{\mathrm{it},t})}{F(\mathbf{a}_{\mathrm{BF},t})}
   = 1 - \frac{F(\mathbf{a}_{\mathrm{it},t})}{F(\mathbf{a}_{\mathrm{BF},t})},
\end{align}
for each trial $t=1,\dots,T$. Aggregating across all trials, we define
\begin{align}
   \mathrm{MRG} &= \frac{1}{T}\sum_{t=1}^T \mathrm{gap}_t, \qquad
   \mathrm{XRG} = \max_{t=1,\dots,T} \mathrm{gap}_t,
\end{align}
where MRG (mean relative gap) indicates the average sub-optimality of the iterative method, while XRG (max relative gap) quantifies the worst-case deviation.
A small $\mathrm{gap}_t \leq \gamma$ implies
\begin{align}
   F(\mathbf{a}_{\mathrm{it},t}) \geq (1-\gamma) \, F(\mathbf{a}_{\mathrm{BF},t}),
\end{align}
i.e., the iterative solution achieves at least $(1-\gamma)$ of the optimal objective value. Hence,
\begin{align}
   \mathbb{E}\{F(\mathbf{a}_{\mathrm{it}}) \} 
   \;\gtrsim\; (1-\mathrm{MRG}) \, \mathbb{E}\{F(\mathbf{x}_{\mathrm{BF}}) \},
\end{align}
demonstrating that the iterative algorithm attains near-optimal performance on average. Furthermore, while brute-force requires exponential complexity $O(G^M)$ for $G$ candidate positions per antenna, the iterative method only requires $O(IM)$ evaluations (with $I$ iterations), which is polynomial in $M$. This sharp reduction in complexity, combined with small MRG/XRG values, justifies the use of the iterative algorithm in practice.

Table~\ref{Brute force} compares the proposed iterative algorithm against the BF benchmark for different numbers of users $K$ and SNR levels. The results show that the iterative algorithm achieves almost identical performance to brute-force in all scenarios. For instance, at $K=3$ and $\mathrm{SNR}=10$ dB, the iterative solution attains $0.1370$ bps/Hz compared to the optimal $0.1385$ bps/Hz, corresponding to an MRG of only $1.1\%$. Similarly, for $K=4$ at the same SNR, the gap is less than $0.1\%$. Even at higher SNR ($20$ dB), where performance differences can become more pronounced, the iterative method remains very close to optimal: the maximum deviation is $3.4\%$ for $K=3$, while for $K=4$ the gap is only $0.2\%$. Overall, the consistently small MRG and XRG values across all cases confirm that the iterative algorithm provides near-optimal performance while avoiding the prohibitive complexity of brute-force search.
\begin{table}[!] 
\centering
\caption{Comparison of iterative algorithm and brute-force benchmark.}
\begin{tabular}{|c|c|c|c|}
\hline
Scenario & $R_\text{BF}$ (bps/Hz) & $R_{\text{it}}$ (bps/Hz) & MRG / XRG \\ \hline
$K=3, \mathrm{SNR=10}$  & 0.1385 & 0.1370 & 0.011/0.011 \\ \hline
$K=4, \mathrm{SNR=10}$  & 0.2618 & 0.2616 & 0.0008/0.0008 \\ \hline
$K=3, \mathrm{SNR=20}$  & 0.9778 & 0.9639 & 0.034/0.034 \\ \hline
$K=4, \mathrm{SNR=20}$  & 1.1591 & 1.1559 & 0.002/0.002 \\ \hline
\end{tabular}\label{Brute force}
\end{table}

\section{The proposed CNN-based Power Allocation}
Following the optimization of antenna placement, we now focus on designing a power allocation mechanism that adapts efficiently to varying user and channel conditions in NOMA systems. Conventional analytical models can characterize these relationships, but the mapping between users’ spatial distribution, channel conditions, and optimal power allocation is inherently nonlinear and environment-dependent.

To address this challenge, we propose a CNN-based learning framework that learns this nonlinear mapping directly from data. By training the CNN on a broad set of channel realizations and corresponding optimal power allocation solutions, the model captures the underlying structural dependencies between input parameters and optimal outputs. Once trained, the CNN can infer near-optimal power allocations for new, unseen network configurations, even under different user densities or antenna settings, without the need for retraining or manual system adjustments.
This data-driven approach enables a unified power control framework that can be readily deployed across diverse NOMA scenarios, combining predictive accuracy with adaptability. The architecture, training procedure, and evaluation of the proposed CNN model are detailed in the following subsections.

\subsection{Architecture of the CNN}

The proposed CNN architecture shown in Fig.~\ref{NetBlockDiagram} begins with an input layer and is followed by a series of convolutional layers, activation functions, and fully connected layers, culminating in a regression layer for output prediction. In a typical CNN-based feature extraction pipeline, convolutional layers apply multiple learnable filters in parallel to capture local patterns in the input data. The output of each convolutional layer is then passed through a nonlinear activation function, producing a set of feature maps. 
Finally, a regression layer is used to produce the desired output data. The details of each layer and the parameters are discussed below.

\subsubsection{Input and Output Layers} \label{input layer}
For generating the input of the proposed framework, we generate an input vector where all the elements of the channel vector, $\mathbf{g} = [g_1, g_2, \cdots, g_K]^T$ are reshaped by the real and imaginary parts as follows
\begin{align}
    \mathfrak{R} &= [\mathfrak{R}(g_1), \cdots, \mathfrak{R}(g_K)] \in \mathbb{R}^{1 \times K}.\nonumber \\
    \mathfrak{J} &= [\mathfrak{J}(g_1),  \cdots, \mathfrak{J}(g_K)]\in \mathbb{R}^{1 \times K}
\end{align}
where $\mathfrak{R}(g)$ and $\mathfrak{J}(g)$ represent the real and imaginary parts of the channel vector, respectively. This 2D-CNN representation preserves both the amplitude–phase relationships
and the user-to-user spatial correlations inherent in PA channels. By applying 2D convolutions across this matrix, the network can jointly capture dependencies between the real and imaginary components as well as across neighboring users. This design leverages the channel’s natural two-dimensional structure, enabling the CNN to learn interference patterns and power-allocation behaviors more effectively.

To enable learning, it is essential to provide the neural networks with paired input data and corresponding ground truth labels (i.e. $(\mathbf{g}, \mathbf{q})$). The performance and accuracy of the trained model depend on the quality of these labels. In this work, the ground truth label vector $\mathbf{q}  \in \mathbb{R}^{K \times 1} $ is obtained by solving the optimization problem defined in \eqref{sm24}, which is associated with each input sample $\mathbf{g} $.

\begin{figure}[!] \label{NetBlockDiagram}
  \includegraphics[width=0.48\textwidth]{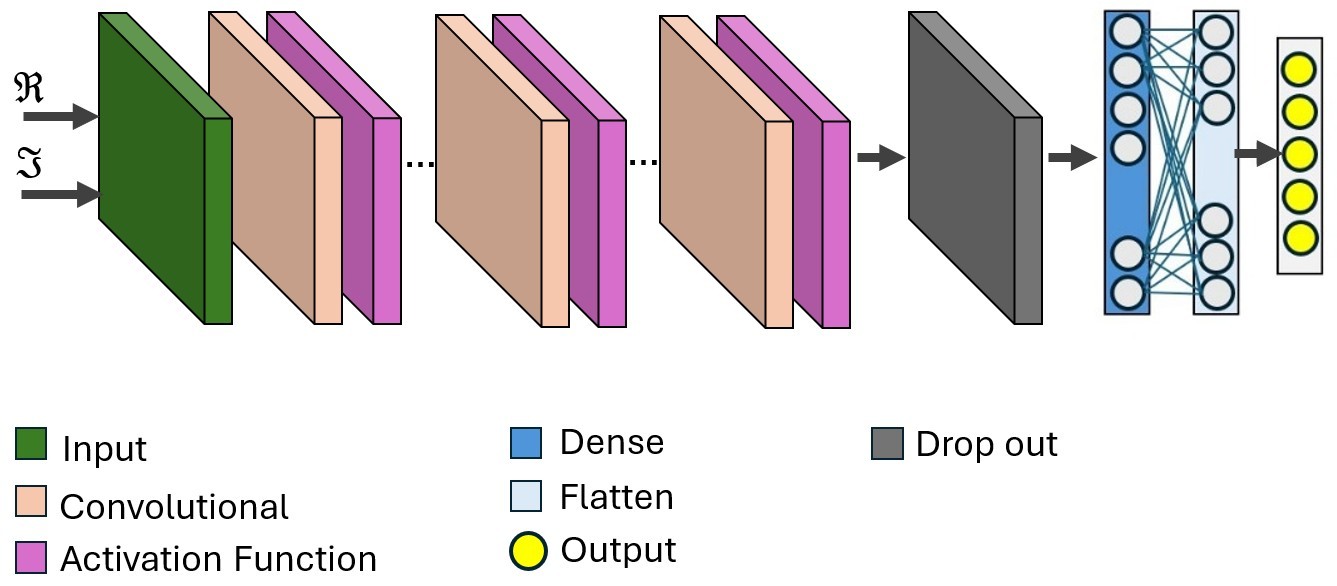}
  \caption{Block diagram of the proposed CNN architecture where $\mathfrak{R}$ and $\mathfrak{J}$ denote the real part and imaginary part of the complex channel vector. }\label{NetBlockDiagram}
\end{figure}

\subsubsection{Activation Function}
In the proposed CNN architecture, the rectified linear unit (ReLU) is employed as the activation function in all convolutional and fully connected layers, except the final regression output layer. The ReLU function introduces non-linearity into the network, enabling it to learn complex representations from the input data. Its computational simplicity and effectiveness in mitigating the vanishing gradient problem make it a widely adopted choice in deep learning models.
The ReLU activation function is defined by $f(x) = \max(0, x)$, 
where $x \in \mathbb{R}$ is the input to the activation function, and $f(x)$ is the output after activation. This function outputs the input directly if it is positive; otherwise, it outputs zero. This sparsity-inducing property helps improve training efficiency and generalization capability.

\textbf{Definition:} \textit{In the context of the convolutional layers, given an intermediate feature map $\mathbf{Z}$ computed via a linear transformation (convolution operation), the activation is applied element-wise as:}
$$
\mathbf{A} = f(\mathbf{Z}) = \max(0, \mathbf{Z})
$$
\textit{where $\mathbf{A}$ denotes the activated output passed to the next layer.}\\
\indent The final regression layer does not employ an activation function to allow the network to produce continuous-valued outputs suitable for regression tasks, such as power allocation coefficients.
By integrating ReLU into the architecture, the proposed CNN benefits from faster convergence during training and enhanced capability to model nonlinear relationships inherent in wireless communication channels.
\subsubsection{Framework description:} 
To learn the nonlinear mapping between the channel features and the optimal power distribution across the pinching antennas, a compact 2D CNN architecture is designed. The input to the network is a matrix of size $K \times 2$ as explained in section~\ref{input layer}. The feature extraction stage of the CNN consists of three consecutive convolutional layers, each followed by a Rectified Linear Unit (ReLU) activation function. The first convolutional layer employs $8$ filters of size $2\times2$ with ``same'' padding to preserve the spatial dimensions, extracting local spatial and spectral dependencies among neighboring users. The second convolutional layer increases the feature depth to $16$ filters with the same $2\times2$ kernel size, enabling the network to capture higher-order correlations between the real and imaginary channel components. Similarly, the third convolutional layer uses $32$ filters of size $2\times2$ to further refine the hierarchical features and encode more complex nonlinear spatial relationships between user and received channel responses. After each convolutional operation, a ReLU activation $f(x)=\max(0,x)$ is applied to introduce nonlinearity, accelerate convergence, and avoid vanishing gradient effects.

Following the convolutional blocks, the extracted feature maps are flattened and passed through a fully connected (dense) layer of size 64. This layer aggregates the learned features from all filters and maps them directly to the predicted power coefficients corresponding to each user. To reduce the risk of overfitting, a dropout layer with a dropout rate of 0.25 is applied. Finally, a regression layer is used as the output layer, which minimizes the mean absolute error (MAE) between the predicted power vector $\hat{\mathbf{q}} = [\hat{q}_1,\hat{q}_2,\ldots,\hat{q}_K]$ and the ground-truth target power vector $\mathbf{q}$. The corresponding loss function is expressed by $\mathcal{L}_{\text{MAE}} = 
\frac{1}{K} \sum_{k=1}^{K} 
\big|\hat{q}_k - q_k\big|$. 
This choice of loss function offers robust convergence and stable gradient updates during training, particularly in the presence of outliers or non-uniform power distributions.



\subsection{Training and Testing}

To ensure the training data reflects realistic PA channel statistics, channel samples are generated by incorporating both large-scale path loss and phase terms arising from waveguide-guided and free-space propagation, as defined in \eqref{waveguide channel}–\eqref{pathloss channel}. This procedure accounts for distance-dependent attenuation, LoS characteristics, and waveguide-induced phase shifts, thereby aligning the dataset with practical PA system behavior. To construct the training dataset, 4,500 input-output sample pairs $(\mathbf{g}, \mathbf{q})$ are synthesized from a Gaussian distribution with zero mean and variance $\sigma^2$. An additional 500 test samples are independently generated from the same distribution to evaluate the model's generalization ability on unseen data. The MAE loss is adopted to directly minimize the deviation between predicted and optimal power allocation coefficients, with early stopping employed to prevent overfitting. Since max–min fairness is enforced through the optimization problem in \eqref{sm24} that generates the ground-truth labels, training with MAE ensures that the CNN predictions remain consistent with fairness-oriented allocations rather than unconstrained sum-rate maximization. This design choice allows the model to approximate fairness-aware solutions while retaining low training complexity.

\begin{figure}[!]
    \centering
    \includegraphics[width=0.46\textwidth]{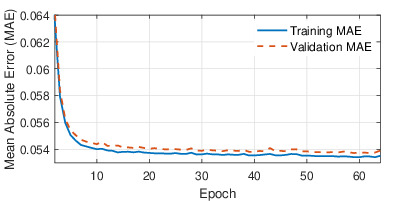}
    \vspace{-0.4em}
    \caption{Learning curve of the proposed model with ADAM optimizer}
    \label{LCSNRNeg10}
\end{figure}

\indent The training process incorporates 5-fold cross-validation to mitigate overfitting further and enhance robustness. This means that our data is randomly divided into 5 groups of equal size. The first fold is treated as a validation set, and the model is fit on the remaining 4 folds. The network is trained over 64 epochs using the adaptive moment estimation (ADAM) optimization algorithm \cite{kingma2014adam}. The learning behavior of the network under a transmit SNR of 10 dB is depicted in Fig.~\ref{LCSNRNeg10}, which demonstrates effective convergence and strong generalization performance.
In all experiments, the learning rate and mini-batch size were set to $10^{-3}$ and 200, respectively. An adaptive exponential learning rate decay was employed, with a decay factor of 0.96 applied every 10 steps. Early stopping was utilized to terminate training when the CNN began to show signs of underfitting or diminishing improvement to ensure optimal performance while maintaining computational efficiency.
\subsection{Discussion on Constraint Enforcement} \label{Discussion}
This section aims to verify that the proposed CNN-based power allocation framework produces outputs that satisfy the required feasibility constraints, i) non-negativity and ii) adherence to the total transmit power budget. The designed CNN effectively learns these constraints during training, enabling it to generate feasible and well-structured power allocation vectors directly from the input channel conditions. The CNN output, denoted by $\hat{\mathbf{q}}\in\mathbb{R}^K$, inherently maintains non-negative values and respects the total power budget, indicating that the network has successfully captured the underlying constraint relationships. To further validate this behavior, the CNN predictions are examined with respect to a simplex scaled by the total transmit power $P$, demonstrating that the outputs consistently fall within or very close to the feasible region. These results confirm the robustness and reliability of the proposed CNN-based power allocation strategy in learning constraint-aware representations without requiring explicit post-processing.\\
We begin by defining the feasible set of power allocations as
\begin{align}
\mathcal{Q} \triangleq 
\big\{ \mathbf{q}\in\mathbb{R}^K \mid 
{q}_i\ge{0},~\sum_{i=1}^K q_i=P \big\}.
\end{align}
If the CNN output $\hat{\mathbf{q}}$ lies within this set, it already satisfies the system constraints and can be directly used. During training, the network is exposed exclusively to feasible targets computed by solving the associated optimization problem, and this repeated exposure enables the CNN to internalize the constraint structure implicitly. Moreover, the use of ReLU activation functions encourages non-negativity naturally. These factors explain the empirical observation that $\hat{\mathbf{q}}$ very often lies within or extremely close to $\mathcal{Q}$ without requiring explicit enforcement.
Nevertheless, in order to guarantee feasibility under all circumstances, the network output is passed through a projection operator that maps any infeasible $\hat{\mathbf{q}}$ onto the nearest feasible point in Euclidean distance. This projection step solves a convex optimization problem of the form
\begin{align}\label{eq:proj_def}
\mathbf{q}^\mathrm{proj} 
= \arg\min_{\mathbf{q}\in\mathcal{Q}} 
\|\mathbf{q}-\hat{\mathbf{q}}\|_2^2.
\end{align}
To understand the structure of this projection and why it can be computed efficiently, we examine the associated Lagrangian, which has a unique solution due to the strict convexity of the objective and the convexity of the simplex $\mathcal{Q}$. To understand the structure of this projection and why it can be computed efficiently, we examine the associated Lagrangian
\begin{align}
 \mathcal{L}(\mathbf{q}, \theta, \nu)=\frac{1}{2}\|\mathbf{q}-\hat{\mathbf{q}}\|_2^2+\theta\Big(\sum_{i=1}^K q_i-P\Big)-\sum_{i=1}^K \nu_i q_i   
\end{align}
where $\theta$ is the multiplier associated with the power constraint and $\nu_i \geq 0$ enforce the non-negativity constraints. From the stationarity condition and complementary slackness:
\begin{align}
    q_i-\hat{q}_i+\theta-\nu_i=0, \qquad \nu_i q_i=0.
\end{align}
It is observed that each coordinate of the solution must either satisfy $q_i=\hat{q}_i-\theta$ whenever it is positive, or be equal to zero whenever $\hat{q}_i-\theta \leq 0$. Thus, the optimal projected allocation has the structure
\begin{align}\label{proj_formula}
q_i^\mathrm{proj} = \max\{\hat{q}_i - \theta,\, 0\},\quad i=1,\dots,K,
\end{align}
revealing that the projection amounts to shifting all components of $\hat{q}$ by the same scalar threshold $\theta$ and clipping at zero where necessary. The entire difficulty lies in identifying the correct value of $\theta$, which must satisfy the power conservation condition
\begin{align}
    \sum_{i=1}^K \max \{ \hat{q}_i - \theta , 0 \} = P.
\end{align}
Because only some components remain active after the shift, the value of the threshold depends on the size of the active set. Sorting the elements of $\hat{q}$ in descending order greatly simplifies this task. Letting $u_1 \geq u_2 \geq \cdots \geq u_K$ denote the sorted values, the active subset corresponds to the largest prefix for which the feasibility condition
\begin{align}
    \rho = \max \Big\{ j : u_j + \frac{1}{j}\Big(P - \sum_{i=1}^j u_i \Big) > 0 \Big\},
\end{align}
holds. Denoting the size of this active set by $\rho$, the threshold is then determined by
\begin{align}
    \theta=\frac{\sum_{i=1}^\rho u_i-P}{\rho},
\end{align}
and substituting this expression into the previous structural formula yields the final projected vector in \eqref{proj_formula}. \\
\indent This derivation confirms that the projection step always yields a feasible solution: any infeasible CNN output is automatically mapped onto the power-allocation simplex, producing a vector that lies exactly within the feasible region and is uniquely closest (in the Euclidean sense) to the original prediction. It also highlights an important feature of the method: since the CNN already learns to respect the constraints, the projection only makes minimal local adjustments. Indeed, numerical evaluations in Section~\ref{Discussion} demonstrate that the differences between $\mathbf
{q}^{\mathrm{proj}}$ and $\hat{\mathbf{q}}$ are typically negligible, meaning that the CNN effectively encodes the feasibility structure as part of its learned representation. The projection, therefore, acts not as a correction mechanism for large errors but as a mathematically principled guarantee of constraint compliance, ensuring robustness and consistency of the proposed learning-based power allocation framework.

\begin{figure}[h]
    \centering
    \includegraphics[width=0.43\textwidth]{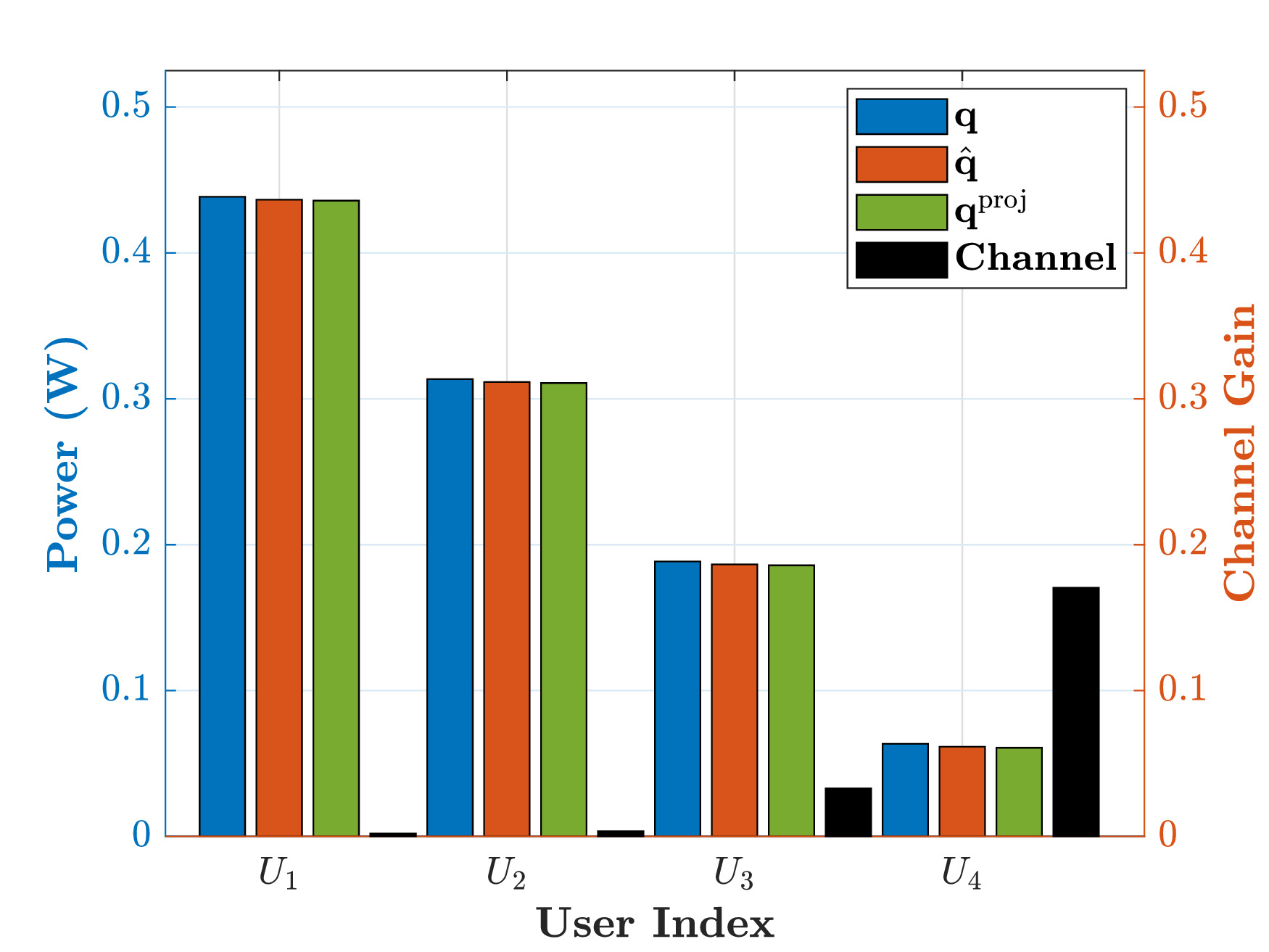}
    \vspace{-0.4em}
    \caption{Projection deviation for allocated power for each user with output of Algorithm 2 (i.e., the ground truth) $\mathbf{q}$, the output of the CNN $\hat{\mathbf{q}}$, and the projected value $\mathbf{q}^\mathrm{proj}$ in Algorithm 3 }
    \label{q-qhat-qpro}
\end{figure}

\begin{algorithm}[!]
\caption{Projection onto the Simplex \cite{duchi2008efficient}}
\label{alg:projection}
\KwIn{CNN output $\hat{\mathbf{q}} \in \mathbb{R}^K$; total transmit power $P$}
\KwOut{Feasible power allocation $\mathbf{q}^\mathrm{proj} \in \mathbb{R}^K$}
\textbf{Parameters:}
\begin{itemize}
  \item $\hat{\mathbf{q}}$: raw CNN output (unconstrained).
  \item $u$: sorted version of $\hat{\mathbf{q}}$ in descending order.
  \item $\rho$: index of the largest feasible subset of $u$.
  \item $\theta$: threshold used to shift the vector back to the simplex.
\end{itemize}
\BlankLine
Sort $\hat{\mathbf{q}}$ into $u$ such that $u_1 \ge u_2 \ge \dots \ge u_K$\;
Find $\rho = \max \left\{ j : u_j + \frac{1}{j}\left(P - \sum_{i=1}^j u_i \right) > 0 \right\}$\;
Compute $\theta = \frac{1}{\rho} \left( \sum_{i=1}^\rho u_i - P \right)$\;
\For{$i=1$ to $K$}{
    $q^\mathrm{proj}_i = \max \{ \hat{q}_i - \theta, \, 0 \}$\;
}
\Return{$\mathbf{q}^\mathrm{proj}$}
\end{algorithm}
As shown in Fig.~\ref{q-qhat-qpro}, the predicted power allocations $\hat{\mathbf{q}}$ generated by the CNN closely follow the optimal ground-truth power $\mathbf{q}$ obtained from Algorithm~2. To ensure constraint satisfaction, the CNN outputs are post-processed using the projection algorithm described in Section~\ref{Discussion}, which maps each prediction onto the feasible simplex defined by the total power constraint, resulting in $\mathbf{q}^{\mathrm{proj}}$. 
The figure demonstrates that $\mathbf{q}^{\mathrm{proj}}$ and $\hat{\mathbf{q}}$ are nearly identical across all users, confirming that the CNN inherently learns to produce feasible allocations without significant correction by the projection step. 
This validates the theoretical claim that if $\hat{\mathbf{q}}\in\mathcal{Q}$, then $\mathbf{q}^{\mathrm{proj}}=\hat{\mathbf{q}}$. 
Furthermore, the variation of power across users reflects the channel-dependent resource allocation behavior: users with weaker channel gains receive higher power to ensure fairness and meet quality-of-service constraints, while those with stronger channels require less power. 
Overall, the negligible difference between $\hat{\mathbf{q}}$ and $\mathbf{q}^{\mathrm{proj}}$ indicates that the CNN not only approximates the optimal allocation accurately but also implicitly learns the underlying constraints of the optimization problem.
This analysis ensures that the proposed learning-based framework maintains both flexibility and constraint compliance, guaranteeing that all power allocations are physically realizable for transmission.

\section{Computational Complexity Analysis}
The primary computational burden in this approach stems from the data generation process necessary for training the CNN, as it requires executing the iterative power allocation algorithm to obtain near-optimal supervision labels. This iterative procedure incurs a complexity of $\mathcal{O}(\log(1/\zeta)K^3)$, where $K$ is the number of users and $\zeta$ denotes the convergence threshold. The computational cost of the CNN during the inference phase is significantly lower compared to traditional optimization-based methods, making it highly suitable for real-time power allocation in NOMA systems. Consider a CNN with $C$ layers, where the $i$-th layer contains $C_i$ neurons. The computational complexity of each layer is dominated by $ C_iC_{i-1}$ real-valued multiplications and additions, corresponding to the matrix operations between consecutive layers. In addition, the total number of activation function evaluations across the network is $\sum_{i=1}^{C} C_i$. This lightweight computational structure enables fast inference, as its complexity primarily depends on the network architecture and remains independent of the number of users once the model has been trained. As a result, the proposed CNN-based approach offers a practical and efficient solution for dynamic power allocation in scenarios with stringent latency and computational constraints. On the other hand, the complexity of the iteration algorithm can be calculated as follows: Let $I$ denote the number of refinement iterations. Each iteration requires $M+1$ evaluations of $\text{SR}(\cdot)$, each invoking an $O(KM)$ channel computation and an inner convex optimization for power allocation. The total complexity is thus approximately $O\big(I(M+1)(KM + C_{\text{inner}})\big)$ where $C_{\text{inner}}$ is the complexity of the convex feasibility solver. Therefore, practical performance depends on limiting $I$, $M$, and $K$.

\section{Simulation results}
\begin{figure}[h]
    \centering
    \includegraphics[width=0.43\textwidth]{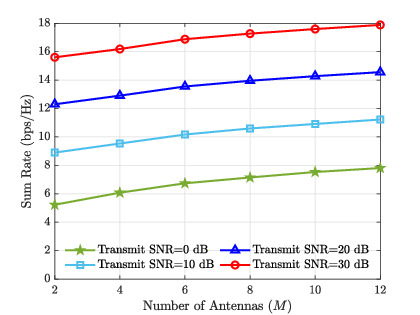}
    \vspace{-0.4em}
    \caption{Sum rate versus the number of antennas for $K=3$ }
    \label{Different M}
\end{figure}

\begin{figure}[!]
    \centering
    \includegraphics[width=0.43\textwidth]{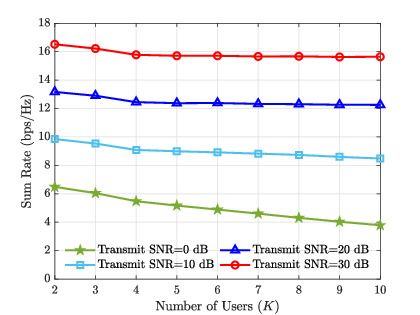}
    \vspace{-0.4em}
    \caption{Sum rate versus the number of users for $M=4$ }
    \label{Different K}
\end{figure}

In this section, computer simulations are used to evaluate the performance of the PA system where \( M=K \) antennas are arranged along a single waveguide. For illustration purposes, the noise power is set as -90 dBm, and $\kappa = 1.4$ \cite{pozar1998microwave}. For the OMA-based downlink communication scenario, where a BS serves $K$ single-antenna users, denoted by $U_k$, which are served in time slot $k$. It is assumed that the $K$ users are uniformly distributed in an area with side length $D_1$ and width $D_2$, as shown in Fig.~\ref{fig:fig1}. The spatial configuration of the antenna elements is defined by coordinates \( \phi_m^\text{pin} \). Users are independently and uniformly distributed within area regions of side length \( D_1 = D_1 = 10\,\text{m} \). The fixed-power NOMA configuration is based on the model in \cite{ding2024flexible}. \\
\indent Fig.~\ref{Different M} illustrates the relationship between the achievable sum rate and the number of antennas $M$ for a fixed number of users $K=3$ under different transmit SNR levels. As observed, the sum rate consistently increases with the number of pinching antennas for all transmit SNR values. This improvement stems from the additional spatial degrees of freedom provided by the PA positions that are dynamically optimized to enhance the received signal strength and reduce inter-user interference. At higher SNR levels, such as 20 dB and 30 dB, the growth in sum rate becomes more prominent due to the enhanced signal alignment and diversity gain achieved by the optimized antenna configuration. It demonstrates that employing multiple PAs significantly boosts system throughput and improves spectral efficiency in multi-user scenarios.\\
\indent Fig.~\ref{Different K} presents the sum rate performance of the proposed pinching antenna system as a function of the number of users 
$K$, for different transmit SNR levels while keeping the number of antennas fixed at $M=6$. It can be seen that the total achievable sum rate decreases gradually as the number of users increases. This behavior occurs because, with more users sharing the same transmission resources, the available power must be distributed among a larger number of users, which increases inter-user interference and reduces the effective signal-to-interference-plus-noise ratio (SINR) per user. At higher SNR levels, such as 20 dB and 30 dB, the performance degradation becomes less severe due to the stronger received signal power and more efficient utilization of the optimized pinching antenna configuration. Conversely, at lower SNR values, the system becomes more sensitive to interference and channel correlation effects, leading to a decline in the sum rate as $K$ grows. 
Overall, the figure demonstrates that the pinching antenna system maintains robust performance even as the number of users increases, showing that the flexible placement and tunable geometry of pinching antennas help mitigate interference and preserve spectral efficiency across a range of user densities.\\
\begin{figure}[h]
\centering
\subfloat[$ K= 2$]{\label{2 user} \includegraphics[width=0.43\textwidth] {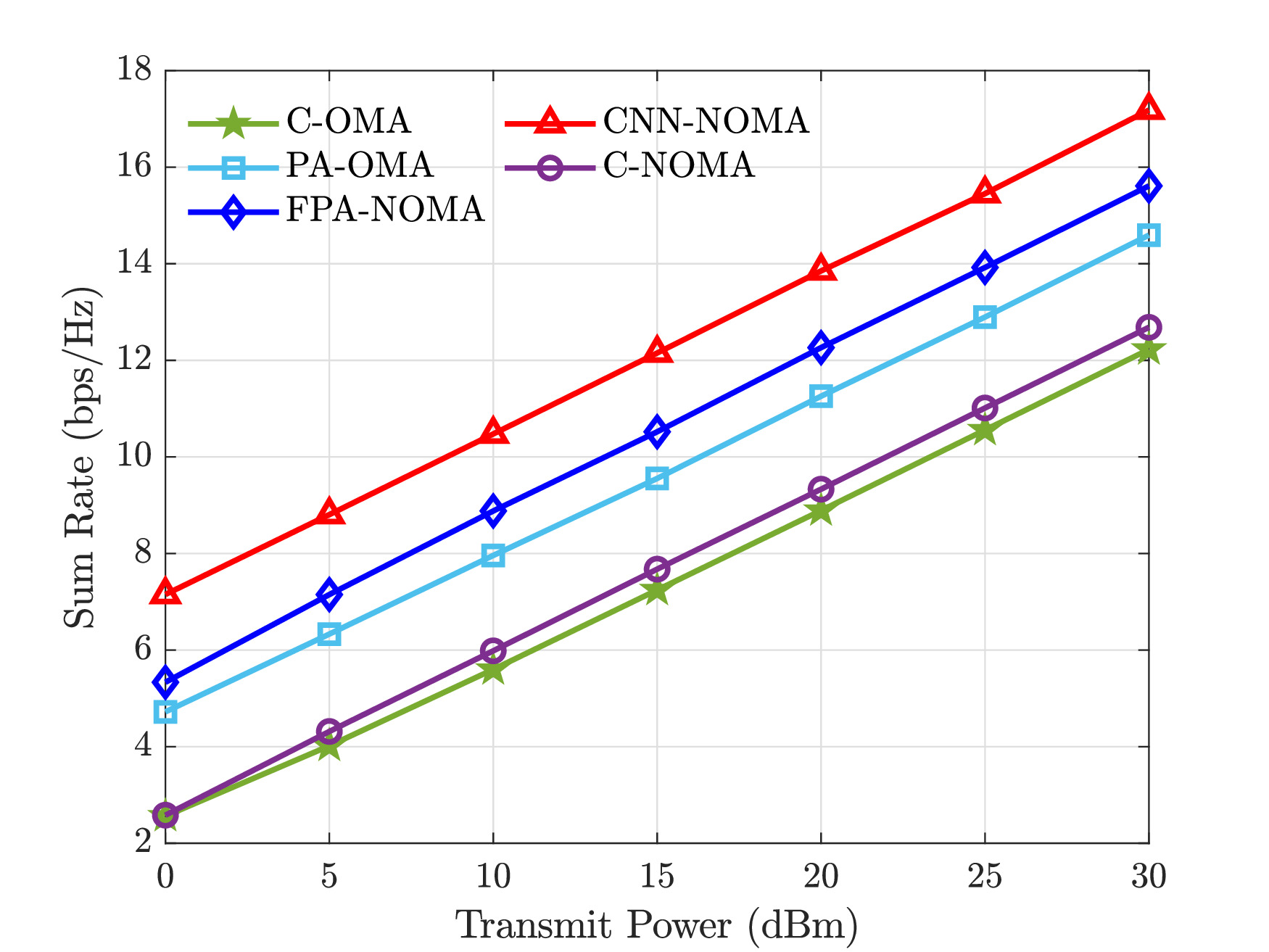}} 
\\[-0.3ex] 
\subfloat[$K = 6 $]{\label{10 user} \includegraphics[width=0.43\textwidth] {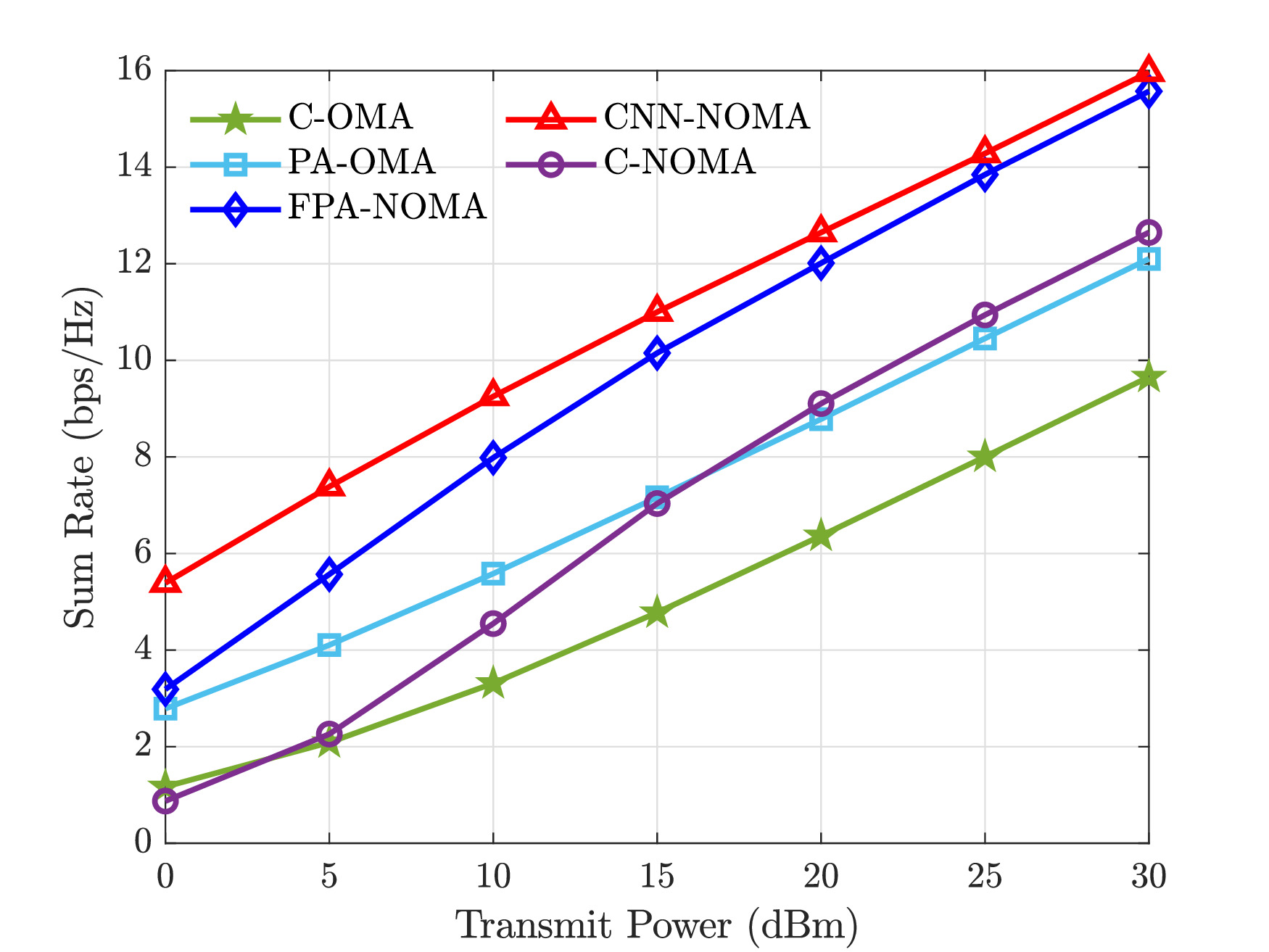}} 
\\[-0.3ex] 
\subfloat[$K = 10 $]{\label{10 user} \includegraphics[width=0.43\textwidth] {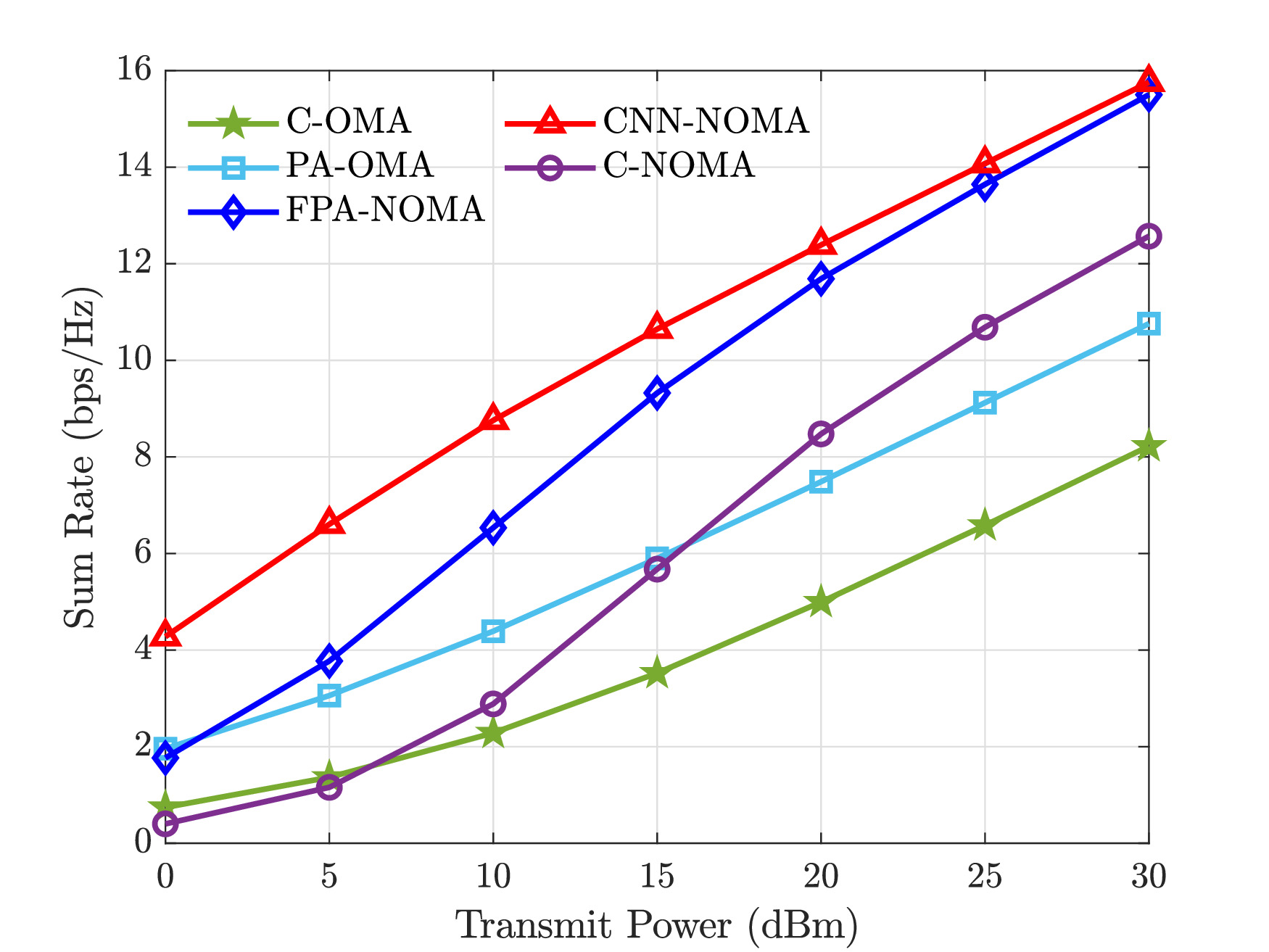}} 
\vspace{-0.4em}
\caption{Sum rate versus the transmit SNR with $M=6$ and $K=2,6,10$.}
\label{NOMA sum rate}
\end{figure}
\indent Fig.~\ref{NOMA sum rate} illustrates the sum rate performance of the tested methods under different numbers of users ($K = 2, 6, 10$) with a fixed number of antennas ($M = 6$). The results are plotted as a function of the transmit power. 
In the proposed CNN-NOMA approach, both the antenna placement along the pinching waveguide and the power allocation among users are optimized iteratively to maximize the overall spectral efficiency. The optimization process jointly adapts the antenna positions to enhance the effective channel gains and allocates transmit power based on the users’ instantaneous channel conditions. 
As observed in all three subplots, CNN-NOMA consistently achieves the highest sum rate compared to other methods, demonstrating the advantage of integrating pinching antenna positioning with power-domain NOMA optimization. The performance gap between CNN-NOMA and conventional schemes increases with higher transmit power, confirming that the proposed optimization framework effectively exploits the spatial degrees of freedom provided by the reconfigurable antenna locations. Additionally, as the number of users $K$ increases, the sum rate improvement becomes more significant, highlighting the scalability of the proposed method in multi-user environments.

In the OMA-based methods, users are served orthogonally in time or frequency, which limits spectral efficiency. Specifically, C-OMA assumes equal power allocation and fixed resource partitioning, while PA-OMA optimizes power allocation among orthogonal users to improve fairness and sum rate. On the other hand, the NOMA-based schemes exploit power-domain multiplexing to serve multiple users simultaneously within the same resource block. FPA-NOMA uses fixed power coefficients, which simplifies implementation but cannot adapt to varying channel conditions. C-NOMA incorporates dynamic power allocation but does not consider the impact of antenna positioning or spatial channel variations.

\begin{figure}[h]
	\centering
		\includegraphics[width=0.43\textwidth]{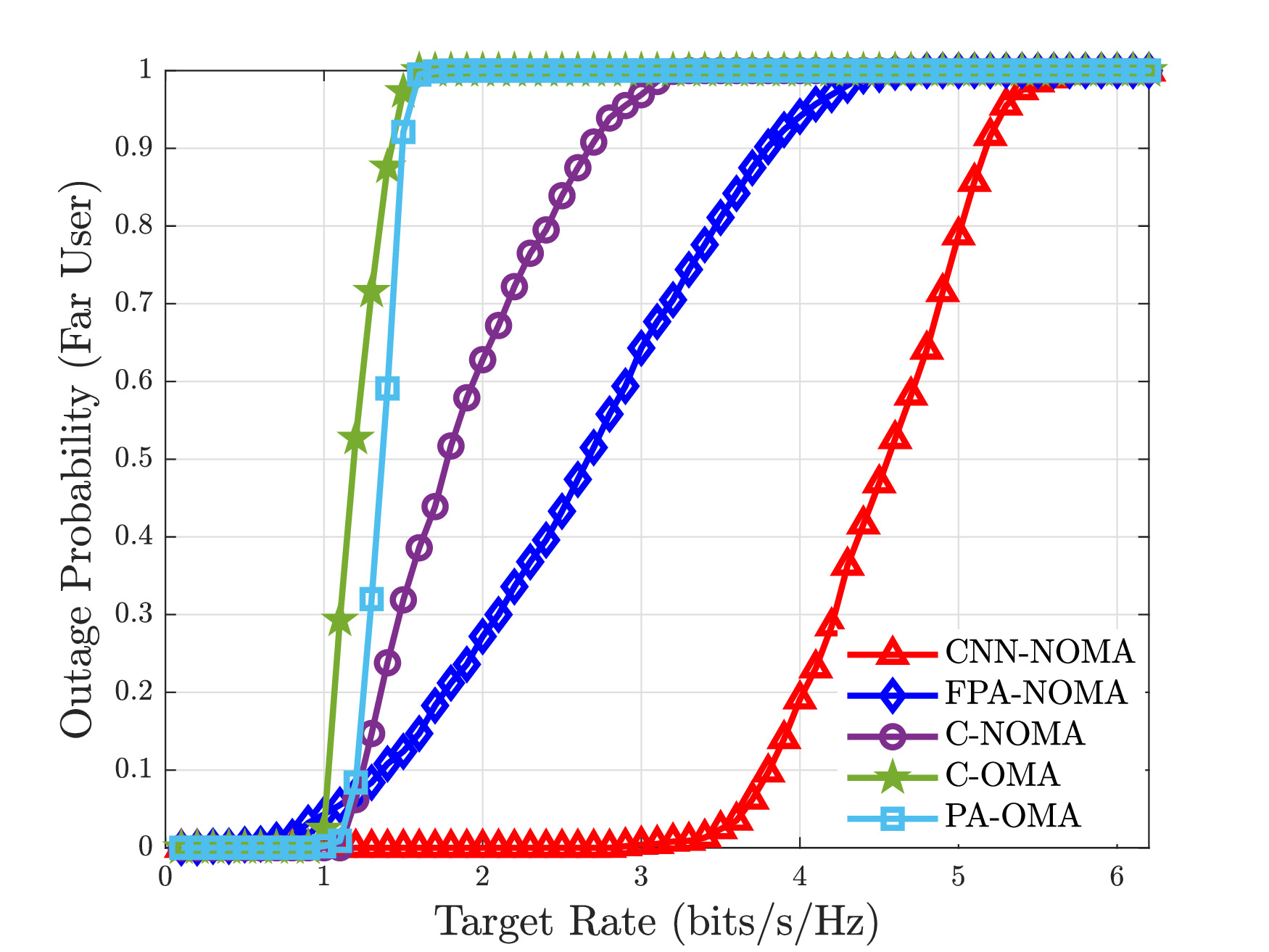}
	\caption{Outage probability of the far user for $M=8$, $K=5$.}
\vspace{-0.4em}
\label{outage}
\end{figure}

In Fig.~\ref{outage}, we analyze the outage probability of far users because it reflects the system’s robustness under worst-case channel conditions and highlights the effectiveness of the proposed algorithm in maintaining reliable service for users with poor channel quality. In this scenario, we assume that there are $K=5$ users and $M=8$ antennas. As shown in the figure, the CNN-NOMA with optimized power allocation and dynamic antenna locations significantly outperforms the other schemes, maintaining near-zero outage probability up to a target rate of approximately 3.8 bps/Hz. This improvement demonstrates the effectiveness of power optimization in exploiting the CNN-based architecture. The FPA-NOMA with fixed power allocation shows a performance drop due to the lack of adaptability to channel conditions, resulting in a steep rise in outage beyond 1.5 bps/Hz. The C-NOMA configuration fails to deliver robust performance at higher target rates, with outage probability exceeding 60\% beyond 2 bps/Hz. Moreover, PA-OMA outperforms C-OMA because the antenna’s active position can be reconfigured along the waveguide to be closer to each user. This reduces path loss, enhances channel gain, and improves signal strength during each user’s transmission slot, leading to a higher achievable sum rate compared to a fixed antenna setup.  These results confirm that incorporating multiple PAs with antenna placement, along with optimized power control, provides substantial reliability gains for the far user in terms of outage performance.

\begin{figure}[h]
	\centering
		\includegraphics[width=0.43\textwidth]{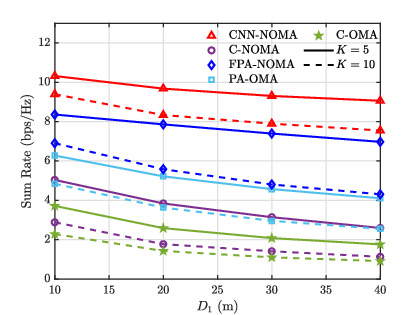}
	\caption{Sum rate versus the different value of the side length $D_1$ with $M=12$, $K=5,10$  }
\vspace{-0.4em}
\label{Diff D1}
\end{figure}

Fig.~\ref{Diff D1} illustrates the variation of the achievable sum rate with respect to the impact of side length $D_1$ (i.e., the
dimension of the user deployment region) for different schemes under two user configurations, $K=5$ and $K=10$, with a fixed number of antennas $M=12$. As the value of $D_1$ increases, the sum rate gradually decreases for all schemes. This trend arises because a larger $D_1$ corresponds to a wider spatial separation between users and the pinching antennas, leading to higher path loss and weaker effective channel gains. Consequently, the received signal power diminishes, reducing the achievable data rate. Among all schemes, the proposed CNN-NOMA consistently achieves the highest sum rate performance across all values of the side length and user configurations. This improvement is attributed to the adaptive power allocation learned by the CNN model and the dynamic location of the antennas close to the users, which effectively captures nonlinear channel interactions and optimally distributes power among users. The C-NOMA and FPA-NOMA methods follow, showing moderate performance, while the PA-OMA and C-OMA schemes exhibit the lowest sum rates due to the lack of efficient spectrum and power reuse. Furthermore, the performance degradation with increasing $D_1$ is more pronounced when the number of users increases from $K=5$ to $K=10$, since the available transmit power is shared among more users, intensifying inter-user interference in the NOMA-based transmission. Overall, the figure confirms the robustness and efficiency of the proposed CNN-assisted power allocation strategy in the pinching antenna system, even under challenging propagation and user distribution scenarios.



\section{Conclusions}
This paper proposed an efficient max-min based power allocation framework for NOMA-enabled pinching antenna systems. First, an iterative gradient-based algorithm was developed to determine the optimal placement of pinching antennas, ensuring maximum sum-rate performance while satisfying practical spacing and boundary constraints. Building on the optimized antenna configuration, a CNN-based power allocation framework was then proposed to learn the nonlinear relationship between system parameters and the optimal power distribution strategy. The trained CNN demonstrates strong generalization to varying numbers of users and antennas, providing a flexible and scalable solution for dynamic wireless environments. Simulation results confirm that the proposed joint design substantially improves spectral efficiency and fairness compared to conventional OMA and non-optimized NOMA schemes, highlighting the effectiveness of integrating learning-based power control with adaptive antenna positioning for next-generation wireless networks.
\section*{Appendix A: Non-convexity of  \eqref{sm24}}
\renewcommand\theequation{A.\arabic{equation}}
\setcounter{equation}{0}

\textit{proof:} From \eqref{channel_g}, the effective channel coefficient is
\begin{align}
g_l = \sqrt{\frac{P}{M}} 
\sum_{m=1}^M 
\frac{\sqrt{\eta}\,
e^{-j 2 \pi \!\left(\tfrac{1}{\lambda}|u_l-\phi_m^{\mathrm{pin}}|
- \tfrac{1}{\lambda_g}|\phi_0^{\mathrm{pin}}-\phi_m^{\mathrm{pin}}|\right)}}{|u_l-\phi_m^{\mathrm{pin}}|},
\end{align}
and we can write
\begin{align}\label{gl2expand}
|g_l|^2 &= \frac{P}{M}\sum_{m=1}^M\sum_{n=1}^M 
\frac{\eta}{|u_l-\phi_m^{\mathrm{pin}}|\,|u_l-\phi_n^{\mathrm{pin}}|} \nonumber \\
&\quad\cdot \exp\Big(-j2\pi\big(\tfrac{1}{\lambda}(|u_l-\phi_m^{\mathrm{pin}}|-|u_l-\phi_n^{\mathrm{pin}}|) \nonumber\\
&\qquad\qquad -\tfrac{1}{\lambda_g}(|\phi_0^{\mathrm{pin}}-\phi_m^{\mathrm{pin}}|-|\phi_0^{\mathrm{pin}}-\phi_n^{\mathrm{pin}}|)\big)\Big).
\end{align}
We prove non-convexity by examining the simplest nontrivial pinching case, \(M=2\). Therefore,\eqref{gl2expand} can be written as
\begin{align}
|g_l|^2
&= \frac{P}{M}\Bigg(
\frac{\eta}{|u_l-\phi_1^{\mathrm{pin}}|^2}
+\frac{\eta}{|u_l-\phi_2^{\mathrm{pin}}|^2} \nonumber\\[-4pt]
&\qquad\qquad
+ \frac{2\eta}{|u_l-\phi_1^{\mathrm{pin}}|\,|u_l-\phi_2^{\mathrm{pin}}|} \cdot \nonumber \\
&\cos\!\Big(2\pi\Big(\tfrac{1}{\lambda}(|u_l-\phi_1^{\mathrm{pin}}|-|u_l-\phi_2^{\mathrm{pin}}|)
\nonumber\\[-4pt]
&\qquad\qquad\qquad - \tfrac{1}{\lambda_g}(|\phi_0^{\mathrm{pin}}-\phi_1^{\mathrm{pin}}|-|\phi_0^{\mathrm{pin}}-\phi_2^{\mathrm{pin}}|)\Big)\Big)\Bigg) \nonumber \\
&= \frac{P\eta}{M}\!\Bigg(\frac{1}{r_1^2}+\frac{1}{r_2^2}
+\frac{2\cos\Delta(\phi_1^{\mathrm{pin}})}{r_1r_2}\Bigg),
\end{align}
with $r_i = u_l-\phi_i^{\mathrm{pin}}$ and 
\begin{align}
\Delta(\phi_1^{\mathrm{pin}})=2\pi\!\left(\tfrac{r_1-r_2}{\lambda}-\tfrac{(\phi_0^{\mathrm{pin}}-\phi_1^{\mathrm{pin}})-(\phi_0^{\mathrm{pin}}-\phi_2^{\mathrm{pin}})}{\lambda_g}\right).
\end{align}
Since $d\Delta/d\phi_1^{\mathrm{pin}}=2\pi(-1/\lambda+1/\lambda_g)=:\Delta'\neq0$, differentiating twice yields
\begin{align}
\frac{d^2 |g_l|^2}{d(\phi_1^{\mathrm{pin}})^2}
=2A\!\left(\frac{3}{r_1^{4}}
+\frac{2\cos\Delta}{r_1^{3}r_2}
-\frac{2\Delta'\sin\Delta}{r_1^{2}r_2}
-\frac{(\Delta')^2\cos\Delta}{r_1r_2}\right),
\end{align}
where $A=\tfrac{P\eta}{M}$. The sign of this expression depends on $\sin\Delta$ and $\cos\Delta$, which vary continuously with geometry. 
To show explicit sign variability, consider the symmetric case $r_1=r_2=R>0$:  
for $\cos\Delta=1$,
\begin{align}
\frac{d^2|g_l|^2}{d(\phi_1^{\mathrm{pin}})^2}
=\frac{2A}{R^4}\!\left(5-(\Delta'R)^2\right),
\end{align}
which is positive when $R^2<5/(\Delta')^2$ and negative otherwise.  
For $\cos\Delta=-1$, it becomes
\begin{align}
\frac{d^2|g_l|^2}{d(\phi_1^{\mathrm{pin}})^2}
=\frac{2A}{R^4}\!\left(1+(\Delta'R)^2\right)>0.
\end{align}
Thus, by varying the antenna geometry or relative phase, the curvature changes sign, proving that $|g_l|^2$ is neither convex nor concave in $\phi_1^{\mathrm{pin}}$.

For the SINR expression
\begin{align}
\mathrm{SINR}_{l,k}(\mathbf q)
=\frac{|g_l|^2 q_k}{|g_l|^2 \sum_{j=k+1}^K q_j + \sigma_{\mathrm{pin}}^2}
=h(|g_l|^2),
\end{align}
where $h(x)=\tfrac{ax}{bx+c}$ with $a,b,c>0$, note that $h'(x)>0$ and $h''(x)<0$ for $x>0$, hence $h$ is increasing and concave.  
Since $x(\phi)=|g_l|^2$ is non-convex, the composite $h(x(\phi))$ inherits this property:
\begin{align}
\frac{d^2}{d\phi^2}h(x(\phi))
=h''(x)\big(x'(\phi)\big)^2+h'(x)x''(\phi),
\end{align}
whose sign varies because $x''(\phi)$ can be both positive and negative. Therefore, the SINR, and any max–min objective involving it, is non-convex with respect to the antenna geometry or phase variables.

\bibliographystyle{IEEEtran}
\bibliography{Ref}

\end{document}